\documentclass[aip,pop,preprint,groupedaddress]{revtex4-1}

\draft % marks overfull lines with a black rule on the right
\usepackage{amsmath,amssymb}
\usepackage{graphicx}
\usepackage{dcolumn}% Align table columns on decimal point
\usepackage{bm}% bold math
\usepackage{color}
\usepackage[caption=false]{subfig}
\usepackage{hyperref}   % use for hypertext links, including those to external documents and URLs
\usepackage[all]{hypcap}
\hypersetup{
    colorlinks=true,
    linkcolor=blue,
    filecolor=magenta,      
    urlcolor=cyan,
    citecolor=red
}
\newcommand{\be}{\begin{equation}}
\newcommand{\ee}{\end{equation}}

\newcommand{\bea}{\begin{eqnarray}}
\newcommand{\eea}{\end{eqnarray}}

\begin{document}
%\title{1D PIC simulation of Buneman instability}
\title{Particle-in-cell simulation of Buneman instability beyond quasilinear saturation}
 \author{Roopendra Singh Rajawat}
\email{rupendra@ipr.res.in}
 \author{Sudip Sengupta}
\affiliation{Institute for Plasma Research, HBNI, Bhat , Gandhinagar - 382428, India }
\date{\today}
%{for revtex  maketitle should be written just here}
%\maketitle 
\begin{abstract} 
Spatio-temporal evolution of Buneman instability has been followed numerically till its quasilinear quenching and beyond, using an in-house developed electrostatic 1D particle-in-cell simulation code. For different initial drift velocities $k_{L}v_{0}/\omega_{pe} \approx 0.1 \, - \, 1$ and for a wide range of electron to ion mass ratios (m/M), 
growth rate obtained from simulation agrees well with the numerical solution of the fourth order dispersion relation. Quasi-linear saturation of Buneman instability occurs when ratio of electrostatic field energy density ($\sum\limits_{k} |E_{k}|^{2}/8{\pi}$) to initial electron drift kinetic energy density ($W_{0} = \frac{1}{2} n_{0}m v^{2}_{0}$) reaches up to a constant value, which as predicted by Hirose [Plasma Physics 20, 481(1978)], is independent of initial electron drift velocity but depends on electron to ion mass ratio m/M as  $\sum\limits_{k} |E_{k}|^{2}/16{\pi}W_{0} \approx (m/M)^{1/3}$. This result stands verified in our simulations. Growth of the instability beyond the first saturation (quasilinear saturation ) till its final saturation [Ishihara et. al., PRL 44, 1404(1980)] follows an algebraic scaling with time. In contrast to the quasilinear saturation, the ratio of final saturated electrostatic field energy density to initial kinetic energy density, is relatively independent of electron to ion mass ratio and is found to depend only on the initial drift velocity. 
Beyond the final saturation, electron phase space holes coupled to large amplitude ion solitary waves, a state known as coupled hole-soliton , are seen in our simulations. The propagation characteristics ( amplitude - speed relation ) of these coherent modes is found to be consistent with the theory of Saeki et. al. [PRL 80, 1224(1998)].
\end{abstract}
% for revtex4 here maketitle should be written
\maketitle 
%\begin{multicols}{2}
\section{Introduction} \label{Introduction}
%\begin{linenumbers}
{Streaming plasmas plays a key role in the generation of shock waves \cite{Bret_pop_2010}, enhances turbulence in tokamaks \cite{Kluiver_pr_1991}, induces anomalous resistivity  \cite{Drake_s_2003, Che_pop_2017} and used in astrophysical scenarios, {\it viz.}, shock surfing accleration \cite{Shimada_pop_2003}, formation of strong double layer \cite{Nsingh_pp_1982, Kaw_pop_2012} , generation of broadband electrostatic noise \cite{Omura_grl_1994} etc. Instabilities\cite{Chen,Hasegawa} associated with streaming plasmas are well known current dissipation mechanism in the presence of external electric field or in the field free collision-less plasma. Being a fundamental current carrying instability, Buneman \cite{Bunemana, Buneman} instability has been the center of attraction for decades. Buneman instability gets excited when relative drift velocity between electrons and ions is sufficiently larger than thermal velocity of electrons.
% and when relative drift is comparable to electron thermal velocity, it becomes ion acoustic instability.
 Buneman wave particle interaction induces scattering of the particles that causes strong parallel heating \cite{Che1}. This novel effect is widely observed/used in electron acceleration \cite{Amano, Dieckmann1, Dieckmann2, Sircombe}, ion acceleration \cite{Yin,Albright} and in inertial electrostatic confinement \cite{Tabaka,Tabak,Bandara} etc.
 
Since the pioneer work of Oscar Buneman \cite{Bunemana,Buneman}; a lot of work has been done to understand linear and nonlinear evolution of Buneman instability in the non-relativistic\cite{Ichimaru,Ionson,Hirose,Ishiharaa,Ishihara, Yoona,Yoon, Pavan, Jain, Shokri, Niknam, Hatami, Che1, Che2, Lampe} and relativistic \cite{Yin,Albright,Haas,Hashemzadeh,Roopendra} regime. Various approaches are attempted by several authors\cite{Ichimaru,Ionson,Hirose} to estimate the saturation value of the Buneman instability; among them Hirose's\cite{Hirose} model successfully predicted that at the quasilinear saturation (or first saturation) the ratio of electrostatic energy density ($\sum\limits_{k} |E_{k}|^{2}/8{\pi}$) to initial kinetic energy density ($W_{0} = (1/2)n_{0}mv^{2}_{0}$) varies with electron to ion mass ratio as $\sim (m/M)^{1/3}$. Ishihara et. al.\cite{Ishiharaa,Ishihara} derived a nonlinear dispersion relation using quasi-linear analysis for  initial delta function distribution (cold beam) for electrons. Ishihara et. al. carried out 1-D kinetic simulation of Buneman instability and compared numerical solution of the nonlinear dispersion relation with simulation results that successfully predicted the breakdown of the linear growth, frequency and growth rate modulation. These authors observed that electron trapping causes the final saturation of the Buneman instability and estimated minimum electrostatic field energy required for quenching of the instability via electron trapping, that scales with initial kinetic energy density as $\sum\limits_{k} |E_{k}|^{2}/16{\pi} \geq 0.11 W_{0}$. 

Yoon\cite{Yoona} has formulated a phase and spatially averaged perturbative nonlinear weak turbulence theory that involves quasi-linear velocity space diffusion and nonlinear wave particle interaction. In the companion paper\cite{Yoon}, Yoon carried out Vlasov simulation of Buneman instability for different electron to ion temperature ratio and compared the simulation results with that derived using weak turbulence theory\cite{Yoona}. Their theory successfully predicted nonlinear development of the Buneman instability qualitatively, when nonlinear scattering term with wave kinetic equation is included . In recently carried out simulation works, Jain et. al.\cite{Jain} and Guo \cite{Guo} have carried out 1-D Vlasov and particle-in-cell simulation respectively. Their simulations show that along with the low frequency Buneman mode and high frequency Langmuir mode; wave modes propagating in the opposite direction of the Buneman wave also gets excited in the nonlinear phase of the instability. Niknam\cite{Niknam} has carried out 1-D particle-in-cell simulation  and reported density steepening at late times as well as dependence of time develpment of electrostatic energy densities with a range of mass ratios. Hashemzadeh \cite{Hashemzadeh_pop_2014,Hashemzadeh_ppcf_2015,Hashemzadeh_physics_2016} has carried out particle-in-cell simulation of Buneman instability for q non-extensive distribution and effect of negative ions on the Buneman instability. There is ample amount of other simulation works dealing with Buneman instability in various applications in space and laboratory plasmas that are too numerous to cite.

Above cited references deal with early nonlinear dynamics or dynamics up to the saturation of Buneman instability. Post saturation dynamics of Buneman instability is still under scanner and to the best in our knowledge very little work has been carried out to understand it. Dynamics after quenching is strongly affected by initial plasma paramter. If initial drift velocity of the electron beam is not much larger than thermal velocity, then initial drift kinetic energy does not dissipate completely \cite{Guo} and some part of it still remains with nonlinear coherent structure. This net drift energy of coherent structures after quenching of Buneman instability may affect interaction between electrons and ions. When initial drift velocity of the electron beam is much larger than thermal velocity of electrons, then initial kinetic energy dissipates completely \cite{Ishiharaa} and a strong interaction between nonlinear coherent structure and surrounding ion may result into formation of coupled hole-soliton \cite{Saeki_jpsj_1991,Saeki1998}. Thus, Buneman instability may decay into ion acoustic wave \cite{Omura} and/or may induce coupled hole-solition \cite{Saeki_jpsj_1991,Saeki1998,Shimada_pop_2003}. Nonetheless, numerous work has been carried out on Buneman instability but quantitative comparison between particle-in-cell simulation and fluid/kinetic model has not been attempted so far.

In this paper, we report quantitative effects of initial drift velocity on the space-time evolution and saturation of linear and nonlinear phase of Buneman instability. We study spatio-temporal evolution of Buneman instability using an in-house developed 1-D electrostatic particle-in-cell code. We have performed four simulation runs for various initial drift velocities $k_{L}v_{0}/{\omega}_{pe} \approx 1, \, 0.5, \, 0.33, \, 0.1$ and observed the effect of initial drift velocity on the growth, quasilinear saturation, final saturation and post saturation dynamics of the Buneman instability. For the sake of completeness in section \ref{theory} we revisit the linear theory of Buneman instability. Section \ref{method of solution} describes a brief description of method of solution. Section \ref{linear} describes evolution of instability upto the quasi-linear saturation. Section \ref{nonlinear} reports evolution of instability upto final saturation and formation of couple hole-soliton after quenching of the instability. We end this report with a summary of our results in section \ref{sum}. 
%%%%%%%%%%%%%%%%%%%%%%%%%%%%%%%%%%%%%%%%%%%%%       
%\clearpage       
%%%%%%%%%%%%%%%%%%%%%%%%%%%%%%%%%%%
\section{Theory} \label{theory}
{Consider a cold electron beam of density $n_{0}$ and velocity $v_{0}$ moving through a homogeneous background of ions of density $n_{0}$. Buneman instability gets excited when initial electron drift velocity is sufficiently larger than electron thermal velocity, {\it i.e.}, $v_{0}/v_{th} \gg 1$. The basic set of fluid equations governing the space-time evolution of Buneman instability in one dimension system can be written as\\
{The continuity equation is}
 \begin{equation} \label{eq:1}
 \frac{\partial n_{s}}{\partial t} + \frac{\partial \left(n_{s}{v}_{s}\right)}{\partial x} = 0,    
  \end{equation}
  The momentum equation is 
  \begin{equation} \label{eq:2}
  \frac{\partial {v}_{s}}{\partial t} + {v}_{s}\frac{\partial \left({v}_{s}\right)}{\partial x} = \pm \frac{eE}{m_{s}},
  \end{equation}
Poisson equation can be written as\\
\begin{equation} \label{eq:3}
\frac{{\partial}E}{\partial x} = 4 \pi e (n_{i} - n_{e}),
\end{equation}
where $s$ stands for species electron/ion, e is charge of electrons and ions and, $v_{s}$, $n_{s}$ and $m_{s}$ are velocity, density and mass of respective species and E is self consistent electric field. Hereinafter we use $m_{e} = m$ and $m_{i} = M$.

For electrons, linearized continuity and momentum equations become\\
\begin{equation} \label{eq:4}
-\iota \omega \delta n_{e} + \iota k n_{0} \delta v_{e} + \iota k v_{0} n_{e} = 0,
\end{equation}
\begin{equation} \label{eq:5}
-\iota \omega \delta v_{e} + \iota k v_{0} \delta v_{e} = -\frac{eE}{m},
\end{equation}
where $\delta n_{e}$ and $\delta v_{e}$ are respectively the perturbed density and velocity. Eliminating $\delta v_{e}$ from equation (\ref{eq:4}) and (\ref{eq:5}), perturbed electron density is\\
\begin{equation} \label{eq:6}
\delta n_{e}=\frac{-\iota e} {m(\omega - kv_{0})^{2}}E.
\end{equation}
Following a similar procedure as above, the linearized perturbed ion density is given by
\begin{equation} \label{eq:9}
\delta n_{i} = \frac{\iota e k n_{0}}{M \omega^{2}}E.
\end{equation}
Now linearized Poisson equation can be written as
\begin{equation} \label{eq:10}
\iota k E = 4 \pi (\delta n_{i} - \delta n_{e}).
\end{equation}
Using equation (\ref{eq:6}),(\ref{eq:9}) and (\ref{eq:10}), we get linear dispersion relation as\\
	 \begin{equation} \label{eq:11}
	  1 = \frac{{\omega}^{2}_{pi}}{{\omega}^{2}} + \frac{{\omega}^{2}_{pe}}{({\omega} - kv_{0})^{2}},
	 \end{equation}
where $k$ is a wave number, $\omega_{pi} = \sqrt{\frac{4 \pi n_{0} e^{2}}{M}}$ and $\omega_{pe} = \sqrt{\frac{4 \pi n_{0} e^{2}}{m}}$ are ion and electron plasma frequencies, respectively.

Using the resonance condition, $kv_{0} \approx \omega_{pe}$ gives\\
\begin{equation} \label{eq:15}
1 \approx \frac{{\omega}^{2}_{pi}}{{\omega}^{2}} + \frac{{\omega}^{2}_{pe}}{({\omega} - {\omega}_{pe})^{2}}.
\end{equation}
Expanding the denominator using $\omega/\omega_{pe} \ll 1$, gives the following relation
\begin{equation} \label{eq:16}
{\omega}^{3}  = - \frac{m}{2M}{\omega}_{pe}^{3},
\end{equation}
which is a cubic equation. \\
Complex roots of cubic equation can be written as\\
	 \begin{equation} \label{eq:18}
 	\Omega = {\omega} + \iota \gamma = (1 \pm \iota \sqrt{3}) \left(\frac{m}{16M} \right)^{(1/3)} {\omega}_{pe},
	 	 	 \end{equation}
where taking only positive sign gives the growth rate of the most unstable mode as\\
	 \begin{equation} \label{eq:19}
 	\gamma = \sqrt{3} \left(\frac{m}{16M} \right)^{(1/3)} {\omega}_{pe}.
	 	 	 \end{equation}	 	 	 
Even though the relative drift velocity between electrons and ions is the key factor which excites the instability, nevertheless, maximum growth rate still turns out to be independent of the initial drift velocity and merely depends on the electron to ion mass ratio.

\section{Method Of Solution} \label{method of solution}
In order to understand the spatio-temporal evolution of Buneman instability beyond the linear stage, we use an in-house developed one dimensional electrostatic particle-in-cell simulation code. The governing equations, {\it viz.}, the particle position and velocity equations and Poisson equation in normalized forms are 
\begin{eqnarray}
\frac{dx}{dt} = v_{s}(x,t)\\
\frac{dv_{s}}{dt} = \pm E(x,t)\\
\frac{\partial E}{\partial x} = (n_{i} - n_{e})
\end{eqnarray}
Then normalization used are $x \rightarrow k_{L}x$, $t \rightarrow \omega_{pe}t$, $v_{s} \rightarrow k_{L}v_{s}/\omega_{pe}$, $E \rightarrow \frac{ek_{L}E}{m{\omega}_{pe}^{2}}$ and $\phi \rightarrow \frac{ek^{2}_{L}\phi}{m{\omega}_{pe}^{2}}$, where $k_{L}$ is the wavenumber corresponding to the longest wavelength supported by the simulation box.
\begin{table}
\caption{\label{tab:tab1}Simulation Parameter.}
\begin{ruledtabular}
\begin{tabular}{lll}
Parameter &	Symbol   &   Value 	\\
\hline \hline
No of grid points			    &	NG 	  	 			&	1024	\\
System Length				    & 	L	 	 			&	$2 \pi$	\\
Time step					    & 	$\Delta t$  		&	$0.0196349 \, {\omega}_{pe}^{-1}$	\\
Grid Spacing				    & 	$k_{L}\Delta x$ 		&	L/NG = 0.006\\
Total no of electron 		    &	$N_{e}$				&	102400\\
Total no of ion	 			    &	$N_{i}$				&	102400\\
Mass ratio					    &	M/m					&	500, 1836, 18360\\
Electron Plasma Frequency 	    &	$\omega_{pe}$ 		&	1		\\
Ion Plasma Frequency 		    &	$\omega_{pi}$		&	$(m/M)^{1/2} \,{\omega}_{pe}^{-1}$		\\
Initial electron drift velocity	& $k_{L}v_{e0}/\omega_{pe}$ &	0.1, 0.33, 0.5, 1.0	\\
Initial ion drift velocity		& $k_{L}v_{i0}/\omega_{pe}$ &  	0.0	\\
Electron thermal velocity	    &	$v_{th,e}/v_{0}$	&	0.003 \\
Ion thermal velocity	        &	$v_{th,i}/v_{0}$	&	0.0
\end{tabular}
\end{ruledtabular}
\end{table}
Parameters used in the numerical experiment of Buneman instability are written in table \ref{tab:tab1}. System length is chosen to be equal to longest mode(L = $2\pi/k_{L}$, $k_{L}$=1) supported by the system. System length is divided in NG equidistant cells, so field quantities electric field, electron/ion density are calculated at the cell center(grid points) and particle quantities like velocities are calculated at the particle positions. Periodic boundary conditions are used that allows only integer mode as k = 1,2,3...512 in the system. A small thermal spread $<[v(0)-\overline{V}(0)]> = 3 \times 10^{-3}$ added to the electron beam to avoid nonphysical cold beam instability \cite{Birdsall}.  Plasma is cold ($v_{0}/v_{th} \approx 1000$) with a very small thermal spread that fulfills necessary condition $v_{drift} \gg v_{thermal}$, so system has favorable condition for excitation of Buneman instability.

In this simulation, we have followed the ion and electron trajectory in the self consistently generated electric field. Initially electrons and ions are placed in phase space with their respective position and velocity. Then for a given ion and electron density, electric field is calculated on the grid points by solving Poisson's equation. Using this electric field, force is calculated on the grid points and then interpolated on particle positions. Further ion and electron momentum equations are solved using this force that yields a new position and velocity. This new particle position is weighted on the grid points to estimate density over the grid points using second order polynomial interpolation. This process is repeated for thousands of time steps.

\section{Results and Discussion} \label{results}

We begin our simulation from an initial state where all the  electrons are flowing as a whole with a single velocity (delta-function velocity distribution) against a homogeneous background of stationary, cold ions. This initial state is unstable to longitudinal perturbations, and as time progresses, small amplitude density (electron and ion density) and velocity oscillations arise from background noise. Since the system is unstable, as the electron beam provides free energy, these small oscillations begin to grow at the expense of the initial beam kinetic energy. In \ref{linear}, we discuss the evolution of the instability till the quasilinear saturation and in \ref{nonlinear} we present the evolution after quasilinear saturation till the final saturation and beyond.

\subsection{Linear growth and quasilinear saturation} \label{linear}

Initially, the growth of the instability is dominated by the most unstable mode and its harmonics; the most unstable mode being given by the resonance condition $k v_{0}/\omega_{pe} \approx 1$. Therefore for electron beam velocities $k_{L}v_{0}/\omega_{pe} \approx \, 1, \, 0.5, \, 0.33, \, 0.1$, the corresponding most unstable modes are respectively given by $k/k_{L} \approx \, 1, \, 2, \, 3, 10$ where the normalizing wave number $k_{L}$ is associated with the longest wavelength that can be supported by the simulation box size.
Fig. \ref{fig:kspace} shows the evolution of electric field amplitude in Fourier space for the mass ratio M/m = 1836 and for the initial electron drift velocity $k_{L}v_{0}/\omega_{pe} \approx 0.33$. For these parameters, the most unstable mode turns out to be $k/k_{L} \approx 3$ whose growth rate is given by $\gamma_{max}/\omega_{pe} \approx 0.054$. Fig. \ref{fig:wtspace} shows the temporal evolution of different Fourier modes for the same set of parameters. The violet, yellow and red line respectively show the growth of the most unstable mode and its first and second harmonic. Most unstable mode grows with the growth rate $\gamma_{max}/\omega_{pe} \approx 0.054$ and its first and second harmonics, which appear later in time, respectively grow with twice and thrice the growth rate of the most unstable mode. Fig. \ref{fig:comp dispersion} shows the growth rate ($\gamma/\omega_{pe}$) as a function of mode number for different initial drift velocities ($k_{L}v_{0}/\omega_{pe} \approx 0.1, 0.2, 0.33$) and for a fixed mass ratio (M/m = 1836). The continuous line curves show the theoretical growth rate as a function of mode number obtained from numerical solution of the linear dispersion relation, i.e. the solution of the fourth degree polynomial (equation \eqref{eq:15}) while the dots show the growth rate obtained from simulations; which show a reasonably good match between fluid theory and simulation. Fig. \ref{fig:comp dispersion}, also shows that the growth rate of most unstable mode (maxima of the curves) is independent of the initial electron drift velocity, and with increasing $k_{L} v_{0}/\omega_{pe}$ the most unstable mode shifts towards shorter wavenumbers; these are in conformity with equation \eqref{eq:19}. In fig. \ref{fig:comp_growth}, we show the dependence of growth rate of the most unstable mode ($\gamma_{max}/\omega_{pe}$) on  the electron to ion mass ratio (m/M) for a fixed initial drift velocity $k_{L}v_{0}/\omega_{pe} \approx 0.33$. The dots represent the simulation points and continuous line is a fit through the points. The linear variation of $\gamma_{max}/\omega_{pe}$ with $(m/M)^{1/3}$ again confirms equation \eqref{eq:19}.

Quasi-linear growth of the instability ceases when exponential growth of the most unstable mode and its harmonics terminate. Fig. \ref{fig:energy_500} and \ref{fig:energy_1836}, respectively show the temporal evolution of the electrostatic field energy density for different initial electron drift velocities $k_{L}v_{0}/\omega_{pe} \approx 0.1, 0.33, 0.5,1$ , for two different mass ratios $M/m = 500,\, 1836$. At quasilinear saturation, time evolution of electrostatic energy density shows a hiccup as shown in inset of fig. \ref{fig:energy_500} and \ref{fig:energy_1836}. This hiccup represents the first saturation (termination of exponential growth) of the Buneman instability. Since the growth rate of the most unstable mode in the linear regime is independent of the initial electron drift velocity,   the ``hiccups'' in electrostatic field energy, for different initial drift velocities appear nearly at the same time. This first saturation occurs when the ratio of electrostatic energy density ($\sum\limits_{k}|E_{k}|^{2}/16 \pi$) to initial drift kinetic energy density ($\frac{1}{2}n_{0}mv^{2}_{0}$) reaches a constant value  $\approx (m/M)^{(1/3)}$ {\it i.e}
\begin{equation} \label{eq:20}
\sum\limits_{k}\frac{|E_{k}|^{2}}{16 \pi W_{0}} \approx \left( \frac{m}{M}\right) ^{(1/3)}
\end{equation}
where $W_{0}$ is the initial drift kinetic energy density of electrons. The reason for quasilinear saturation of the instability at this low value of the ratio, is because of the narrow width(FWHM) of the growth rate $(\gamma/\omega_{pe})$ vs mode number ($k v_{0}/\omega_{pe}$) curve around the resonance point $kv_{0} \approx \omega_{pe}$ (see fig. \ref{fig:comp dispersion}). This figure shows a drastic drop in the growth rate of the instability for any  small change in the electron drift velocity. When change in drift velocity ($\sim k\Delta v_{0}/\omega_{pe}$) becomes comparable to FWHM ($\sim \Delta(k v_{0}/\omega_{pe}$) of the $\gamma/\omega_{pe}$ vs $k v_{0}/\omega_{pe}$ curve {\it i.e.}($k \Delta v_{0} /\omega_{pe} \approx \Delta(kv_{0}/\omega_{pe})$) then exponential growth of the instability terminates \cite{Hirose}.
Based on this argument and a quasilinear calculation, Hirose et. al. \cite{Hirose} showed that at the first saturation, electrostatic field energy density scales linearly with the initial electron drift kinetic energy density with a slope which depends on the electron to ion mass ratio as $(m/M)^{1/3}$ (equation \eqref{eq:20} above). This result is verified in our simulation as shown in fig. \ref{fig:comp_saturation}, where we have plotted the electrostatic field energy density at the first saturation point vs. initial drift kinetic energy density for different mass ratios 500, 1836 and 18360. The linear variation is in conformity with Hirose's scaling \cite{Hirose}. To the best of our knowledge, this is the first verification of Hirose's scaling using a PIC code.

\subsection{Beyond quasilinear saturation: Formation of Coupled hole solitons} \label{nonlinear}
Termination of quasi linear growth does not imply complete saturation of the instability. Beyond this point the instability evolves with algebraic growth up to the final saturation \cite{Ishiharaa}. 
This algebraic growth stage (time between quasilinear saturation and final saturation) decreases with the decreasing ion to electron mass ratio as shown in fig \ref{fig:energy k1 diff m_M} and \ref{fig:energy k3 diff m_M} where we have plotted the temporal evolution of electrostatic field energy density for $k_{L}v_{0}/\omega_{pe} \approx 0.33$ and $k_{L}v_{0}/\omega_{pe} \approx 1$ respectively, for different mass ratios.
As mentioned earlier, the resonant mode ({\it i.e.} the most unstable mode) and its harmonics govern the evolution of the instability up to the quasi-linear saturation. Beyond the quasilinear saturation,
the evolution of the instability is governed by the rapid growth of the non-resonant modes (see fig. \ref{fig:wtspace}). Evolution of the instability in this regime has been studied by several authors \cite{Hatami}  
who have predicted steepening of electron density profile at late times ({\it i.e.} beyond quasilinear saturation). Figure \ref{fig:ewb} and \ref{fig:iwb} respectively show the time development of electron and ion density profiles at different time steps. Both electron and ion density show small oscillations growing out of background noise; these oscillation eventually steepen and gain large amplitude at late times. 

When the wave potential becomes large enough, some electrons are trapped in this self consistently generated nonlinear wave potential well; these trapped particle population generate a counter streaming population of electrons in the plasma (see figure \ref{fig:electron phase space}). This counter streaming population excites electron-electron two stream instability that leads to the formation of holes in electron phase space.
When large number of electrons are trapped in the wave potential well, the instability saturates abruptly. After completion of trapping, instability is quenched and potential shows sudden phase reversal\cite{Ishiharaa} at the time of trapping. Fig. \ref{potential} shows the evolution of the potential profile beyond the quasilinear saturation upto the final saturation ($\omega_{pe} t / 2 \pi \sim 45 - 55$) for $k_{L} v_{0}/\omega_{pe} \sim 0.33 $ and $M/m = 1836$. Phase reversal of potential is clearly seen at $\omega_{pe} t /2 \pi \sim 53$.
Around the same time the electron phase space plots (see figure \ref{fig:electron phase space}) ) also show enhanced trapping (phase space holes). Based on the argument that the final saturation of the instability is caused by electron trapping, Ishihara\cite{Ishiharaa} et. al. calculated the ratio of electrostatic field energy density to initial electron drift kinetic energy density at the final saturation point and showed that $\sum\limits_{k} \frac{|E_{k}|^{2}}{16 \pi W_{0}} \geq 0.11$. Thus in contrast to quasilinear saturation, this ratio is independent of mass ratio (equation \eqref{eq:20}). Our simulations show, that this ratio is not very sensitive to the mass ratio but depends on the initial electron drift velocity. For the mass ratio M/m = 1836, the ratio of electrostatic field energy density to initial electron drift kinetic energy density at the final saturation varies with initial drift velocities as
\begin{equation} \label{eq:28}
\sum\limits_{k} \frac{|E_{k}|^{2}}{16 \pi W_{0}} \approx 0.11 (k_{L}v_{0}/\omega_{pe} = 0.1) \sim 0.18(k_{L}v_{0}/\omega_{pe} = 1)
\end{equation}
This is in conformity with Ishihara's\cite{Ishihara} inequality. Equation \eqref{eq:28} shows that the field energy required for complete trapping depends on initial drift velocity and increases with increasing initial drift velocity. We have performed simulations with wide range of initial drift velocities and mass ratios ( see figures (\ref{fig:energy_500} and \ref{fig:energy_1836}) which respectively show final saturation level for two different initial drift velocities ($k_{L}v_{0}/\omega_{pe} \approx 0.33\,\,\, \& \,\,\,1.0$), for different mass ratios); and in each case it is found that the ratio of electrostatic field energy density at the final saturation to the initial electron drift kinetic energy density follows Ishihara \cite{Ishiharaa} inequality, {\it i.e.}, ($\sum\limits_{k} \frac{|E_{k}|^{2}}{16 \pi W_{0}} \geq 0.11$). 
  
Figure (\ref{fig:electron phase space}) shows snapshots of electron phase space at different times. As mentioned above, around $\omega_{pe}t / 2 \pi \sim 55$ phase space holes are seen in the electron fluid, the number of holes being equal to the most unstable wavenumber ($k/k_{L} = 3$ for $k_{L}v_{0}/\omega_{pe} \approx 0.33$). At this time {\it i.e.} $\omega_{pe}t/2\pi \sim 55 $ the time of final saturation, the mean electron drift velocity nearly goes to zero. The electron phase space holes are thus almost stationary, resulting in strong interaction with the surrounding ions. This strong interaction of the electron phase space holes with the surrounding ions exhibits very interesting dynamics involving both electrons and ions.
To begin with, the positive potential associated with an electron phase space hole starts reflecting the surrounding ions causing compression in the ion fluid on both sides of the electron hole. This compression induces ion pulses close to the edges of the electron hole which in turn pulls electrons from the edges resulting in disruption of the hole itself. As a consequence, each electron hole (mother hole) is elongated and gets divided into two holes (daughter holes; see time frames between $\omega_{pe}t/2\pi \sim 65 - 70$ in Figure (\ref{fig:electron phase space})). The resulting daughter holes which are accompanied by ion pulses start propagating in directions opposite to each other. Each of these new coherent structure thus formed, is a combination of an electron hole and an ion pulse. Below we identify them as coupled hole-soliton (CHS) as described by Saeki at. al. \cite{Saeki_jpsj_1991,Saeki1998}. For a better visualization of the entire process of breaking of electron phase holes into daughter holes ultimately leading to the formation of coupled hole-solitons, we present the electron phase space, the ion phase space and the associated potential profile, at various time steps for a different initial electron drift velocity ($k_{L}v_{0}/\omega_{pe} \sim 0.5$; see figure \ref{fig:chs}). 

Figure (\ref{fig:chs}) shows evolution of electron phase space and ion phase space along with the potential profile at different instances, for $k_{L}v_{0}/\omega_{pe} \sim 0.5$ and $M/m = 1836$. At $\omega_{pe}t/2{\pi} \approx 63.0$ the electron phase space shows two holes corresponding to the most unstable wave number, which in this case is $k/k_{L} = 2$. These phase space holes are nearly stationary. As time progresses, the dynamics described in the previous paragraph is seen, {\it i.e.} each hole interacts with the surrounding ions, becomes elongated and eventually breaks into two holes which start propagating in opposite directions (see time frames between $\omega_{pe}t/2\pi = 63 \,\, - \,\, 67.6$ in figure (\ref{fig:chs})). The associated potential profile also evolves starting from two peaks at $\omega_{pe}t/2 \pi \sim 63.0$ (corresponding to mother holes) to four peaks at $\omega_{pe}t/2 \pi \sim 67.6$ (corresponding to daughter holes). As mentioned above, each daughter hole is accompanied by an ion pulse and the resultant coherent structures propagate in opposite directions (see time frames between $\omega_{pe}t/2\pi = 67.6 \,\, - \,\, 69.4$ in figure (\ref{fig:chs})). We now compare the relation between the measured speed ( Mach number M ) and the associated potential maximum ($\phi_{max}$) for a daughter hole having phase space area ($S$) with the theoretical relation amongst the same quantities for coupled hole solitons as proposed by Saeki et. al. \cite{Saeki1998}. According to the model proposed by Saeki et. al. \cite{Saeki1998} the phase space area ($S$) of a coupled hole-soliton is related to the associated potential maximum $\phi_{max}$ and its speed (Mach no. $\mathcal{M}$), through the integral 
\begin{equation} \label{eq:29}
S = 4 \int^{\phi_{max}}_{W^{2}_{0}/2} (\frac{-W_{0}^{2} + 2 \phi}{-2V(\phi,\mathcal{M},W_{0},\alpha)})^{1/2} d \phi
\end{equation}
where $S$ is normalized to $(k_{L}\lambda_{D})^{2}$ and $\phi_{max}$ is the normalized maximum potential $e\phi_{max}/T_{e}$. $\alpha^{2}$ is the electron to ion mass ratio ($m/M$) and $W_{0}$ is a parameter which is related to $\phi_{max}$, $\mathcal{M}$ and $\alpha$ through the equation $V(\phi_{max}, \mathcal{M}, W_{0}, \alpha) = 0$, where $V(\phi,\mathcal{M},W_{0},\alpha)$ is the Sagdeev potential which is given by the expression
\begin{equation} \label{eq:30}
\begin{split}
V(\phi,\mathcal{M},W_{0},\alpha) = -\frac{1}{6}\{[(1 - \alpha \mathcal{M})^{2} + 2 \phi]^{3/2} + [(1 + \alpha \mathcal{M})^{2} + 2 \phi]^{3/2} - (1 - \alpha \mathcal{M})^{3} \\ - (1 + \alpha \mathcal{M})^{3}\} + \frac{1}{3} \theta(-W_{0}^{2} + 2 \phi) [-W^{2}_{0} + 2 \phi]^{3/2} + \mathcal{M} [\mathcal{M} - (\mathcal{M}^{2}-2 \phi)^{1/2}]
\end{split}
\end{equation}  
Saeki's \cite{Saeki1998} model for coupled hole-solitons is based on water bag distribution for electrons where the velocity distribution at the position of the hole vanishes. The electron velocity distribution function measured around the holes (shown in red and blue in figure (\ref{fig:water bag phase space})), is shown in figure (\ref{fig:water bag distribution}); it shows a reasonable approximation to the theoretical distribution. The measured phase space area $S$ is similar for the red and blue holes and is around $\sim 1.9$. The respective Mach numbers and the potential maximum are $\mathcal{M} \approx 2.01 \,\,\, , \phi_{max} \approx 0.47 $ and $\mathcal{M} \approx 3.1 \,\,\, ,  \phi_{max} \approx 0.342 $. These points (shown in red and blue dots) lie very well on the continuous $\phi_{max} - \mathcal{M}$ curve generated for $S \approx 1.9$ using Saeki's \cite{Saeki1998} theory (equation \eqref{eq:29} and \eqref{eq:30} ). The black dot shown in the same figure is for another coupled hole soliton ($S \approx 3.4$) which is excited using a different set of initial conditions ( $k_{L}v_{0}/\omega_{pe} \sim 1$ and $M/m = 1836$); thus our simulation results show good agreement with the theory of coupled hole solitons.

After final saturation electrostatic field energy density  decreases sharply (see figure (\ref{fig:energy_1836}) which is plotted for $M/m = 1836$ with different initial electron drift velocities ) and exhibits oscillatory behaviour with a frequency which is approximately twice the ion plasma frequency. Decrease in electrostatic field energy is accompanied by stretching of phase space holes, formation of coupled hole solitons ( as decribed above) and finally detrapping of electrons. This trapping and detrapping of electrons results in heating of electrons at late times $\omega_{pe}t/2\pi \sim 100$ through the process of sepatrix crossing, as discussed in Che et. al., \cite{Che2}. At around $\omega_{pe}t/2{\pi} \approx 100$ electron phase space holes coalesce away. Figure (\ref{fig:electron distribution function}) shows the spatially averaged electron distribution function at different times which clearly show broadening of distribution function at late times.   
\section{Summary}
\label{sum}
In this paper, we have studied spatio-temporal evolution of Buneman instability using an in-house developed 1-D particle-in-cell simulation code. Quasilinear (or first saturation) occurs when the electrostatic energy density becomes $\sim (m/M)^{1/3}$ times the initial drift kinetic energy density, {\it i.e.} $\sum\limits_{k}|E_{k}|^{2}/16 \pi \approx (m/M)^{(1/3)} W_{0}$; the ratio of electrostatic field energy density to the initial drift kinetic energy density at the quasilinear saturation point is independent of the initial drift velocity. Further, electron trapping and nonlinear mode coupling leads to the final saturation of the instability. In contrast to quasilinear saturation, at the final saturation the ratio of electrostatic field energy density to initial kinetic energy density depends on the initial drift velocity but is independent of the mass ratio. The above mentioned ratio follows the inequality suggested by Ishihara et. al. \cite{Ishiharaa}, {\it i.e.}, $\sum\limits_{k} \frac{|E_{k}|^{2}}{8 \pi W_{0}} \geq 0.11$. To the best of our knowledge, this is the first verification of Hirose's \cite{Hirose} and Ishihara's \cite{Ishiharaa, Ishihara} results using a PIC code.

After final quenching of Buneman instability, strong interaction between electron phase space holes and surrounding ions is observed; this interaction breaks the electron phase space holes into two oppositely propagating holes each attached with an ion pulse. These oppositely propagating coherent structures have been identified as coupled hole-solitons using the theory of Saeki et. al. \cite{Saeki1998}. These coupled hole solitons eventually coalesce away finally generating a broadened electron velocity distribution function.
\bibliography{bune}
%%%%%%%%%%%%%%%%%%%%%%%%%%%%%%%%%%%%%%%%%%%%%%%%%%%%%%%%%%%%%%%%%%%%%%%%%%%%%%%%%%%%%%%%%%%%%%%%%%%%%%%%%%%%
\clearpage
%%%%%%%%%%%%%%%%%%%%%%%%%%%%%%%%%%%%%%%%%%%%%%%%%%%%%%%%%%%%%%%%%%%%%%%%%%%%%%%%%%%%%%%%%%%%%%%%%%%%%%%%%%%%
\section{figures}
%\begin{figure}[h]
%\centering
%\includegraphics[scale=0.8]{graph/dispersion_curve_threshold1.pdf}
%\caption{Dispersion curve $\gamma({kv})$ for the Buneman instability that describes thershold of the velocty}
%\label{fig:dispersion}
%\end{figure}
%
\begin{figure}[h]
\centering
\includegraphics[scale=0.8]{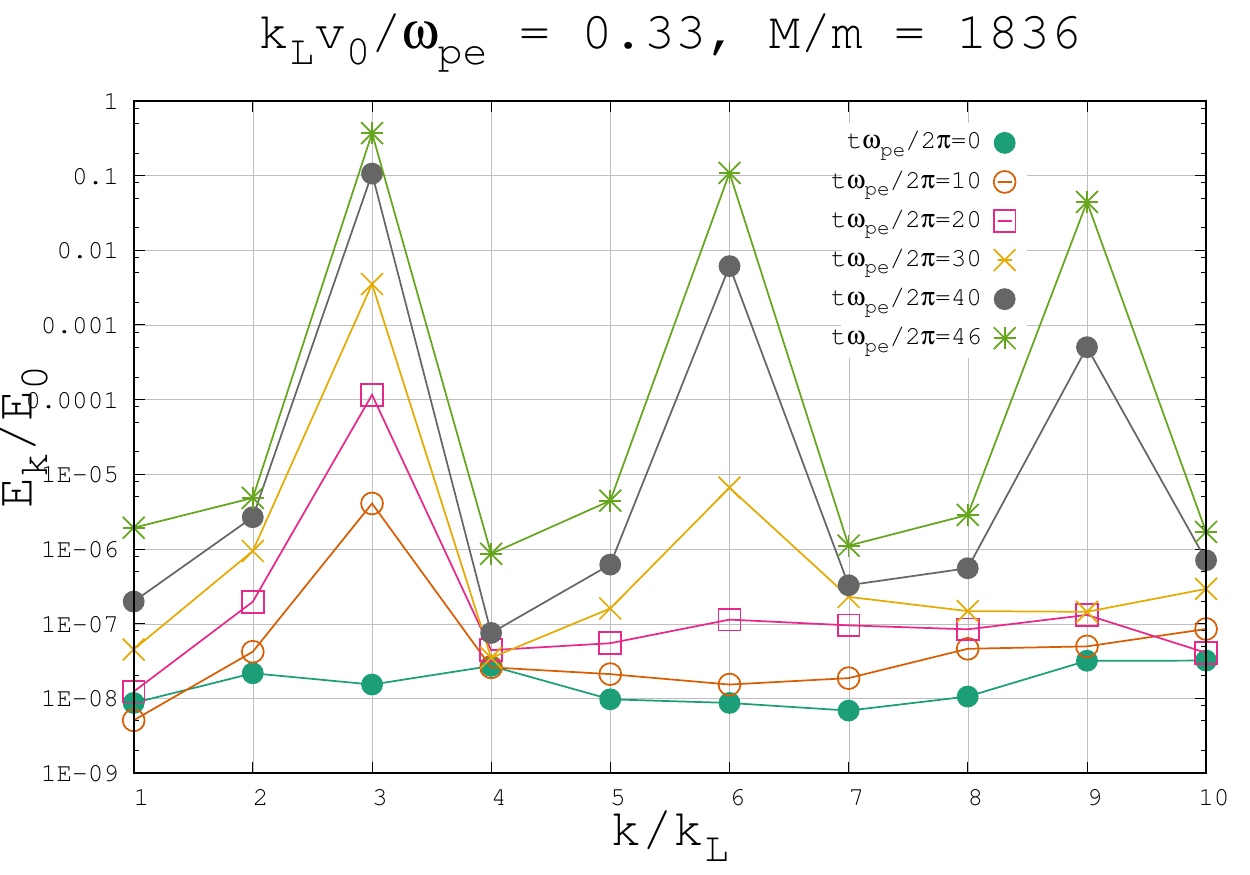}
\caption{Evolution of electric field amplitude in Fourier space for $k_{L}v_{0}/\omega_{pe} \approx 0.33$ and M/m = 1836.}
\label{fig:kspace}
\end{figure}
\begin{figure}
\centering
\includegraphics[scale=0.8]{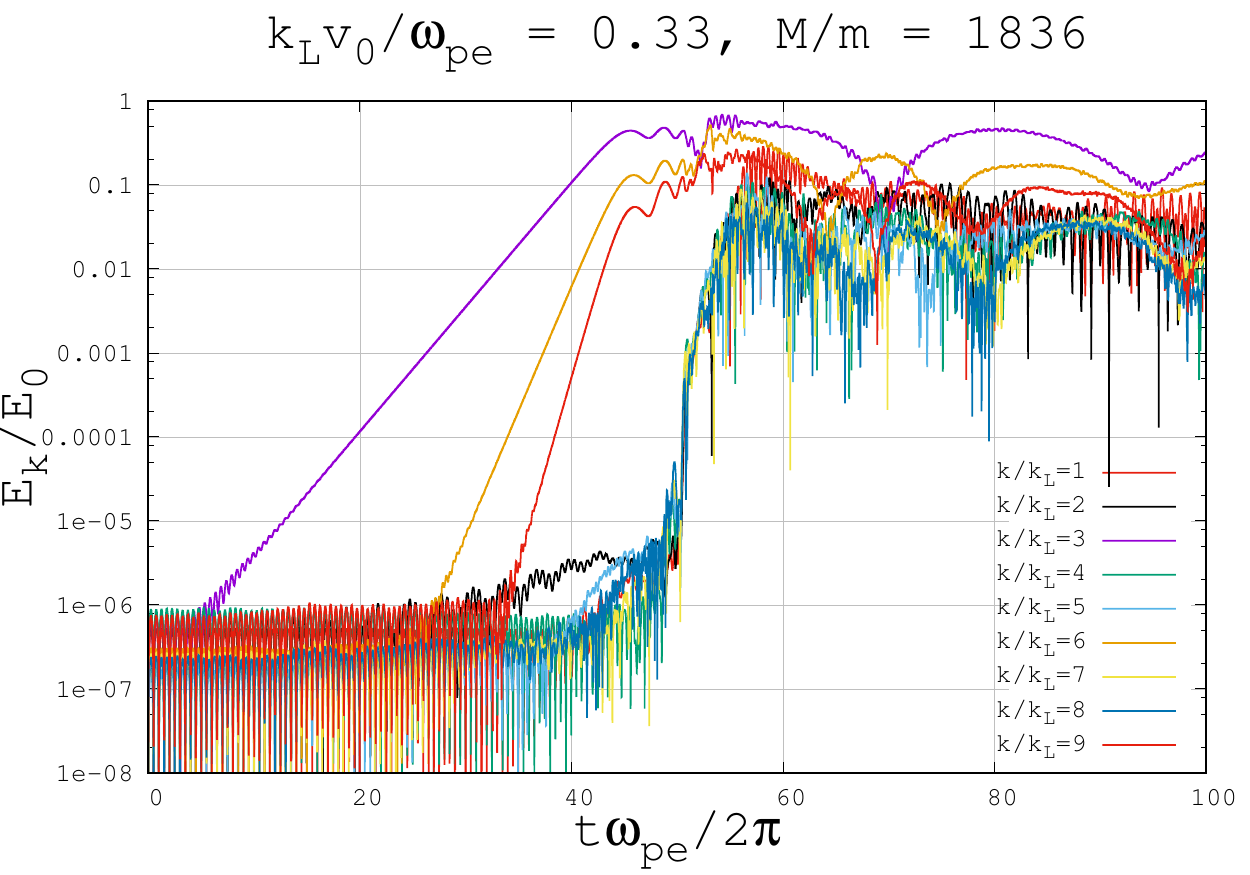}
\caption{Temporal evolution of Fourier modes for  $k_{L}v_{0}/\omega_{pe} \approx 0.33$ and M/m = 1836}
\label{fig:wtspace}
\end{figure}
\begin{figure}
\centering
\includegraphics[scale=0.8]{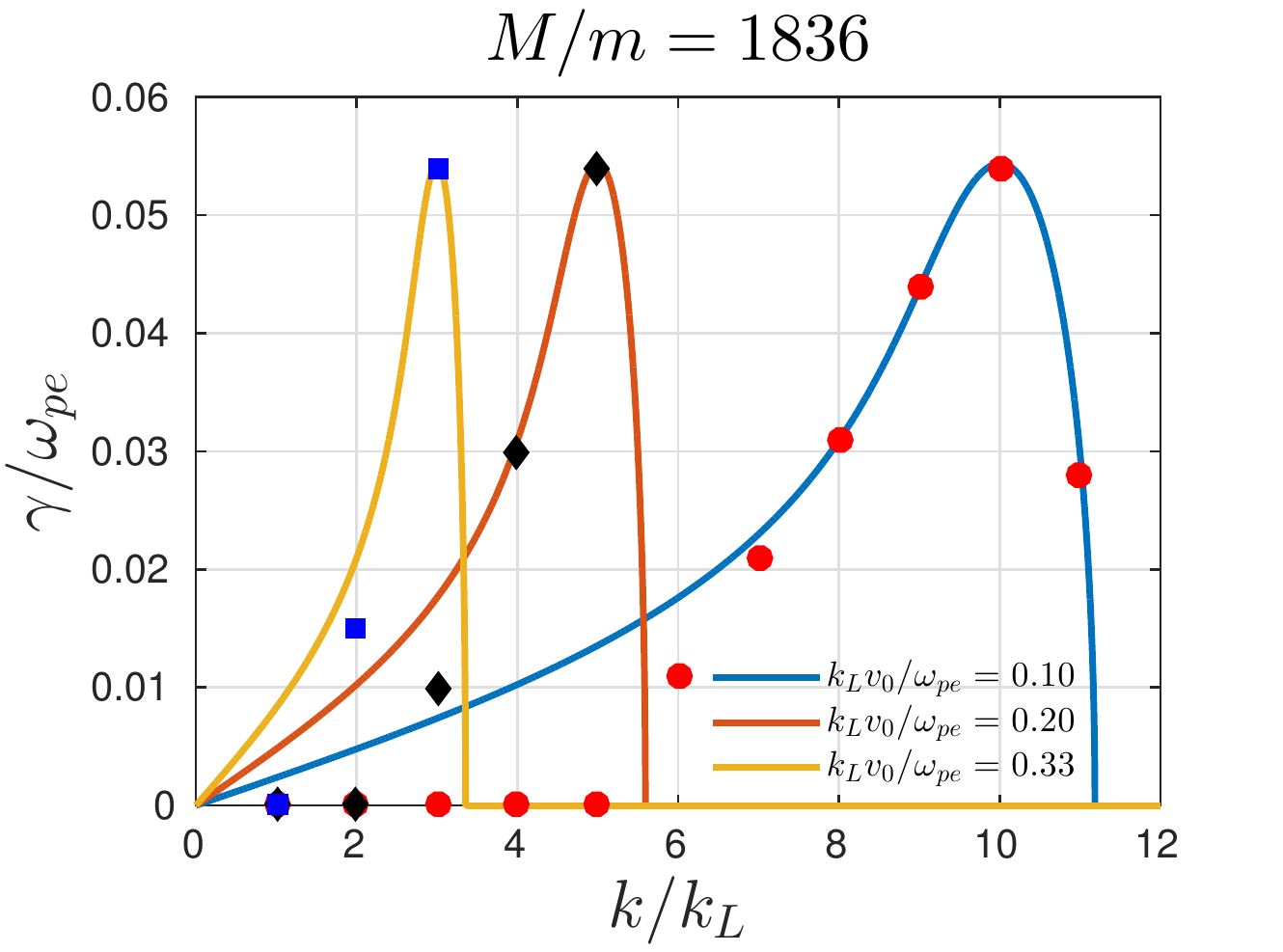}
\caption{Comparison between theoretical growth rate (points) and growth rate obtained from the simulation (continuous line) as a function of mode number ($k/k_{L}$) for the initial drift velocities $k_{L}v_{0}/\omega_{pe} \approx 0.1, \, 0.2, \, 0.33$ and mass ratio M/m = 1836.}
\label{fig:comp dispersion}
\end{figure}
\begin{figure}
\centering
\includegraphics[scale=0.8]{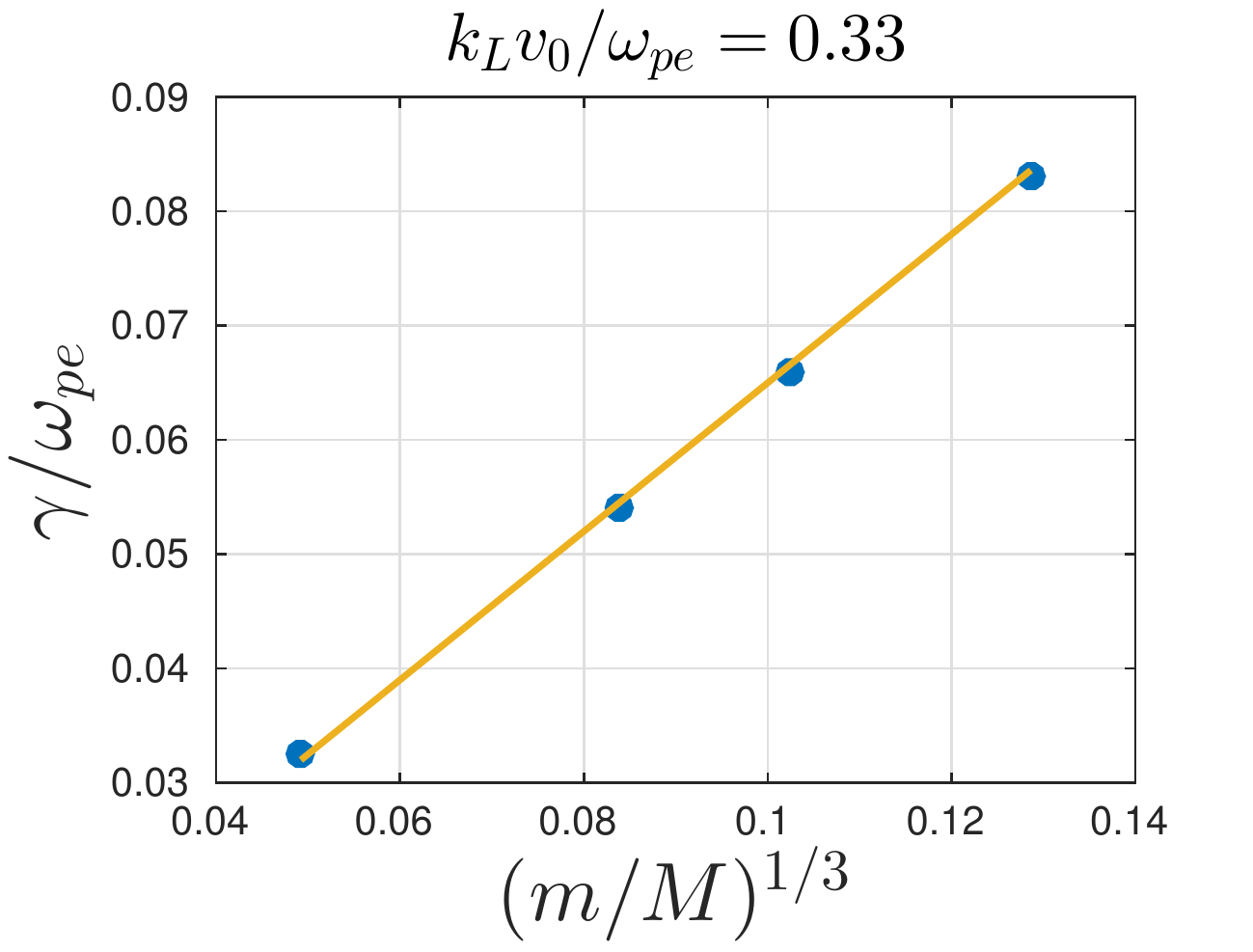}
\caption{Comparison of growth rate of the most unstable mode for the initial drift velocity $k_{L}v_{0}/\omega_{pe} \approx 0.33$ and different mass ratios.}
\label{fig:comp_growth}
\end{figure}
\begin{figure}
\centering
\includegraphics[scale=0.8]{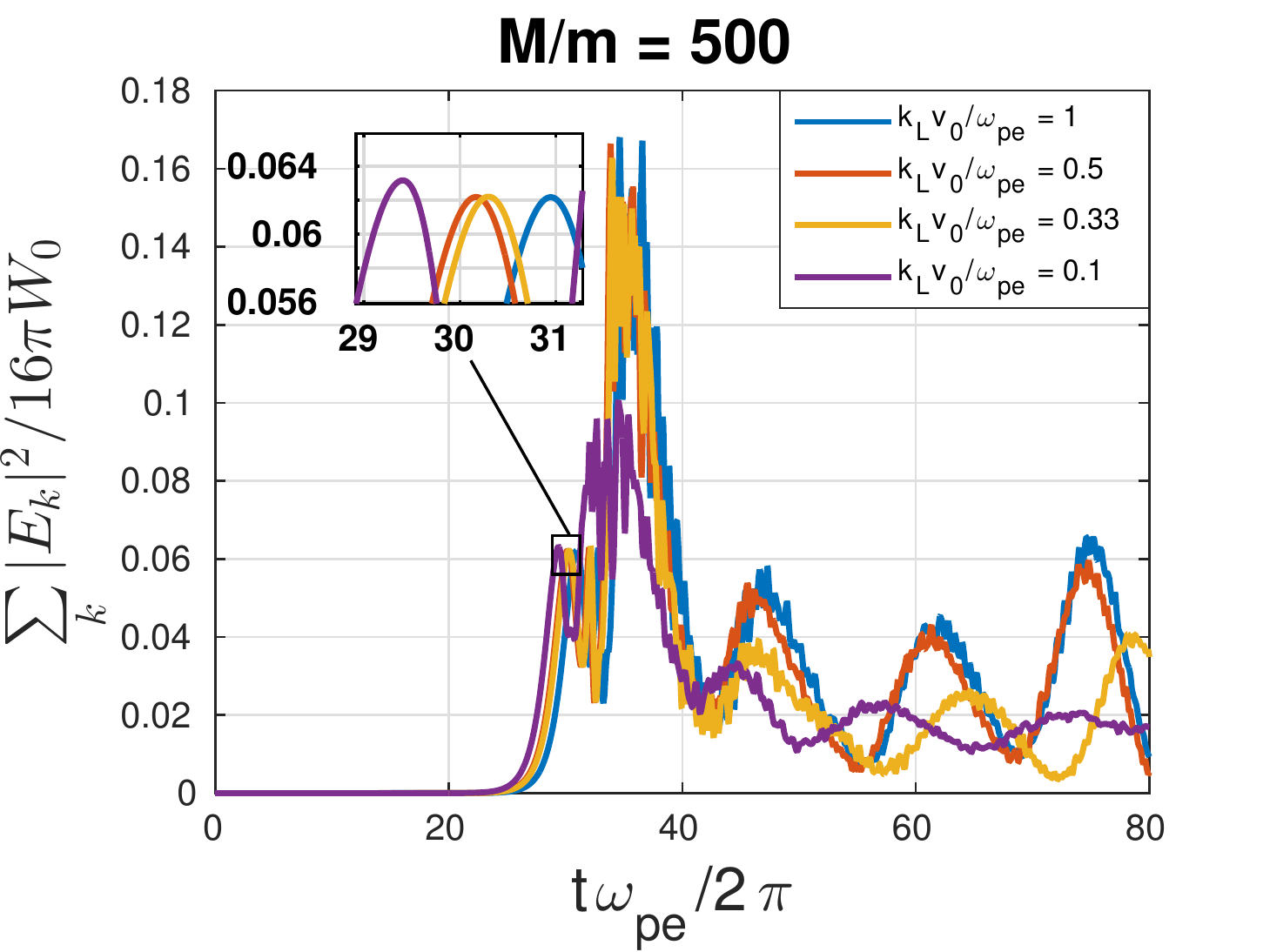}
\caption{Temporal evolution of ratio of electrostatic energy density to different initial drift kinetic energy density for the mass ratio M/m = 500.}
\label{fig:energy_500}
\end{figure}
\begin{figure}
\centering
\includegraphics[scale=0.8]{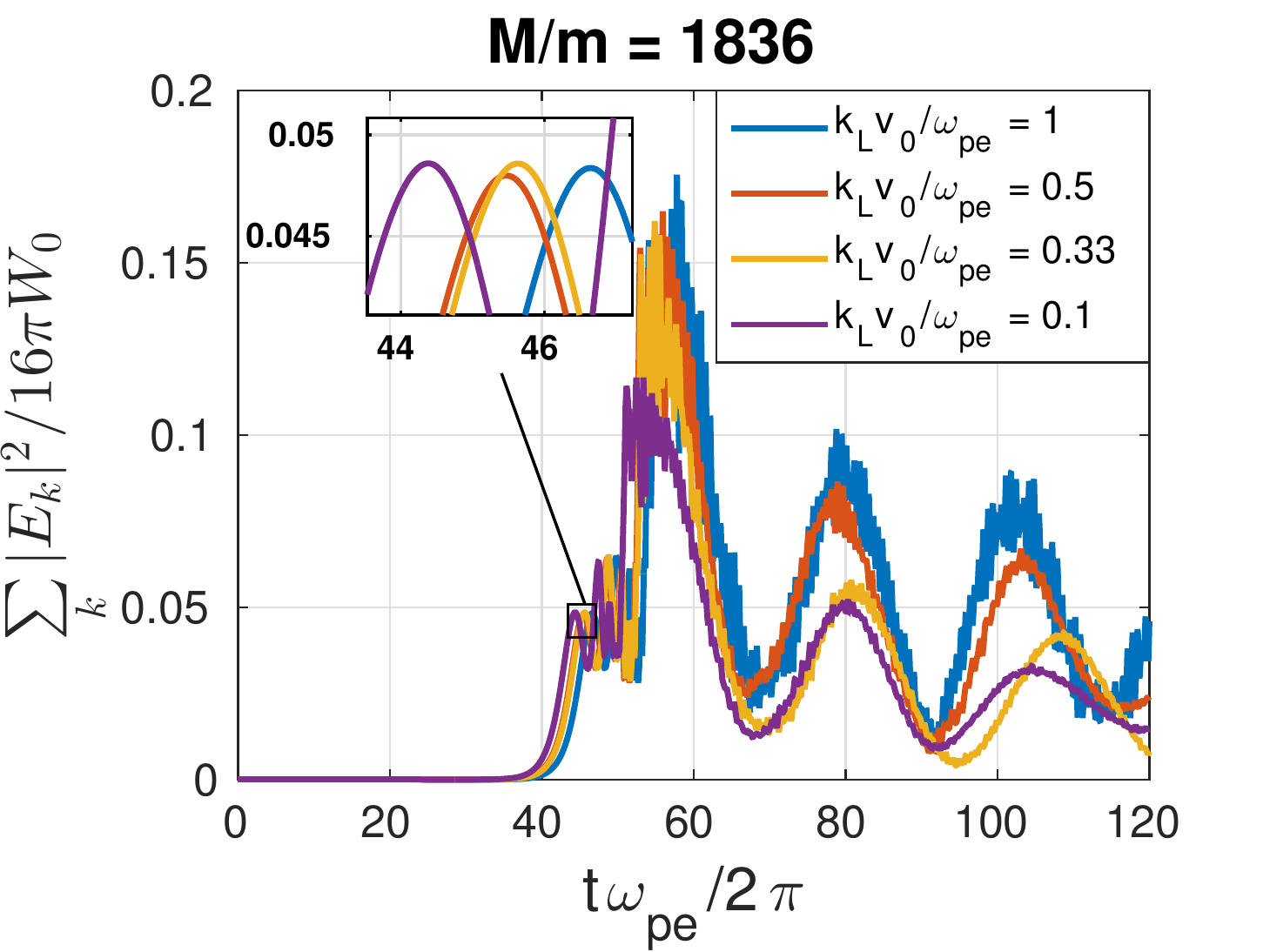}
\caption{Temporal evolution of ratio of electrostatic energy density to different initial drift kinetic energy density for the mass ratio M/m = 1836.}
	\label{fig:energy_1836}
\end{figure}
\begin{figure}
\centering
\includegraphics[scale=0.8]{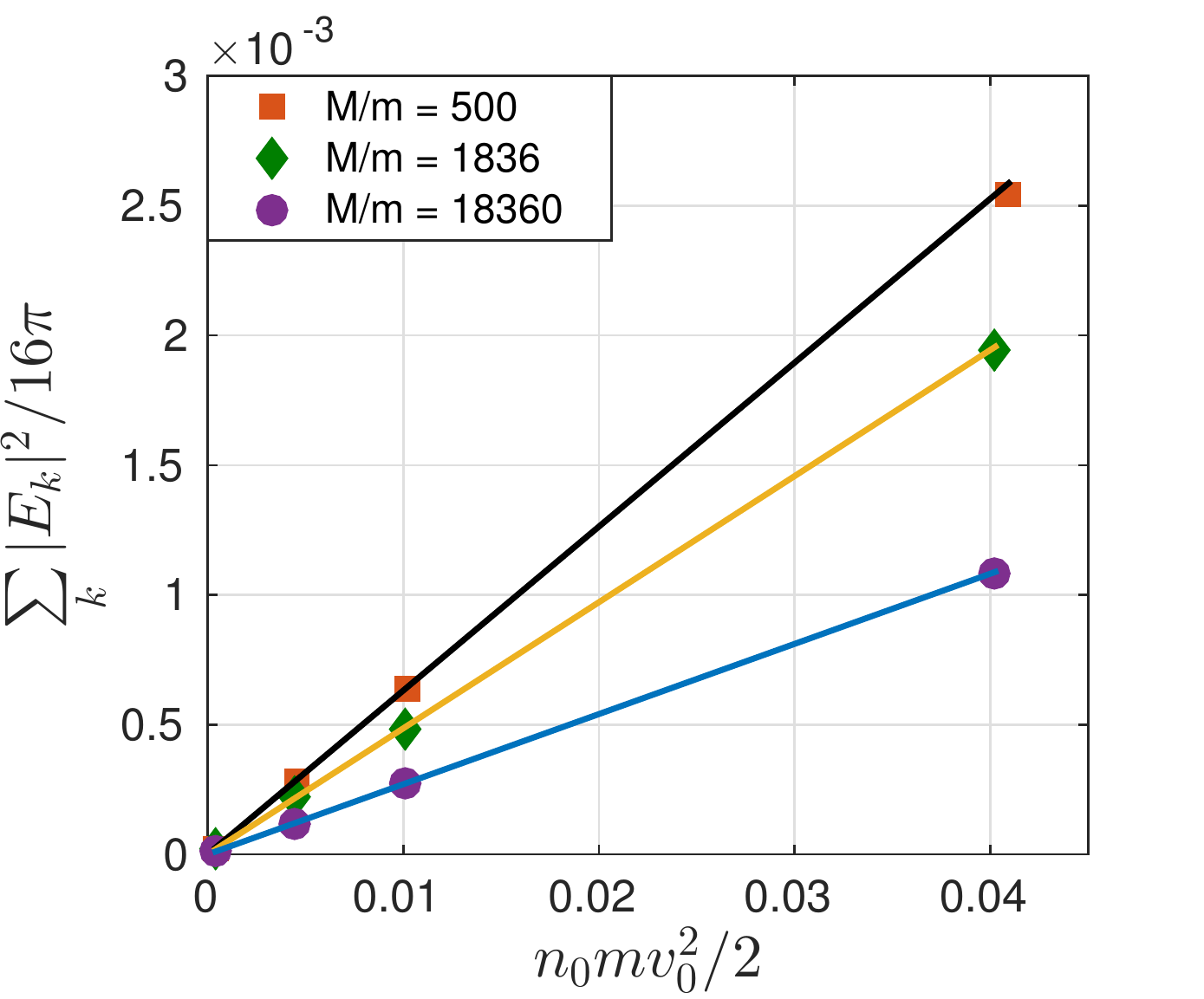}
\caption{Figure shows variation of electrostatic energy density with different initial kinetic energy density for the mass ratios M/m = 500, 1836, 18360.}
\label{fig:comp_saturation}
\end{figure}
\begin{figure}
\centering
\subfloat[]{ \label{fig:energy k3 diff m_M}
\includegraphics[scale=.45]{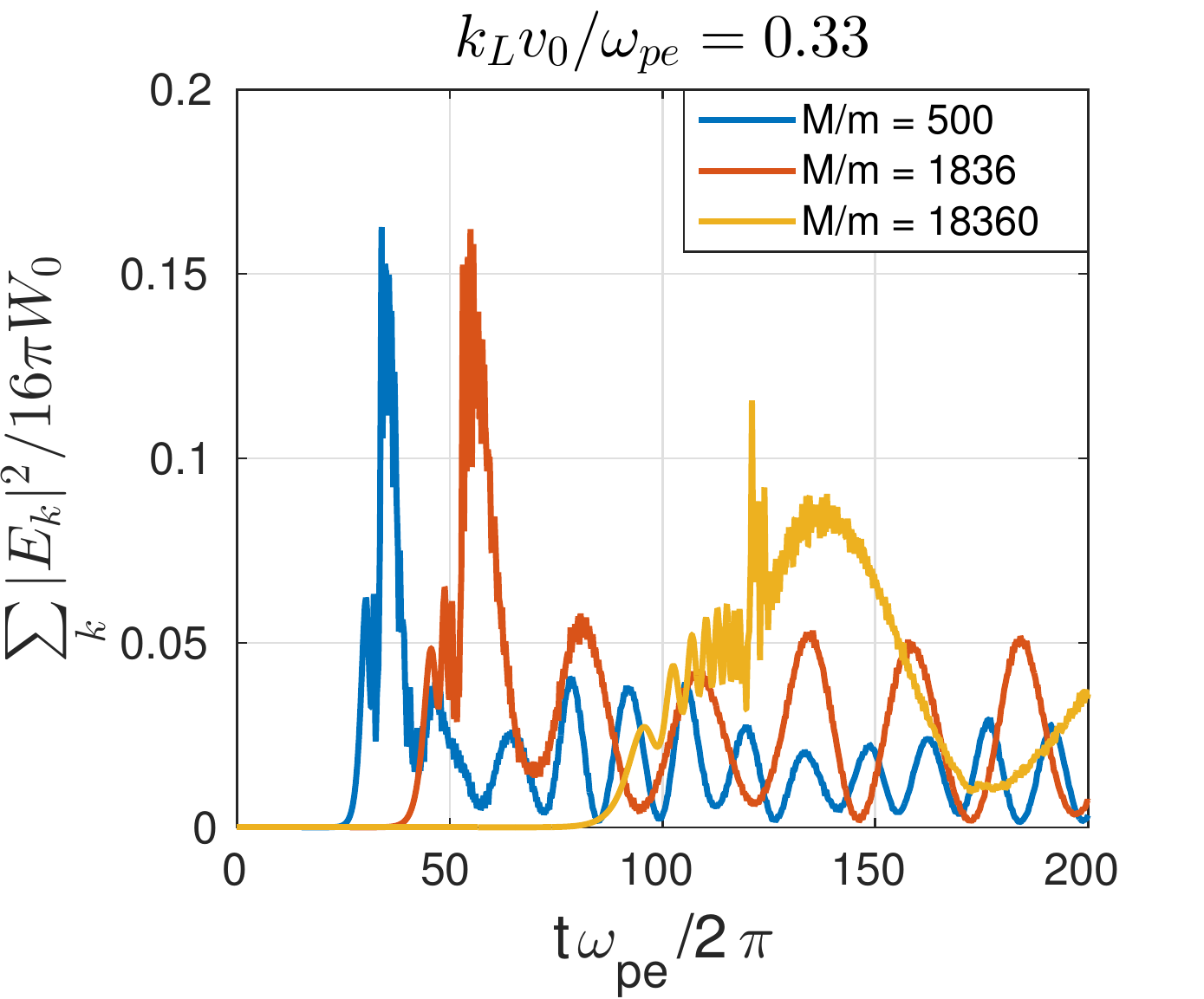}
}
\subfloat[]{ \label{fig:energy k1 diff m_M}
\includegraphics[scale=.45]{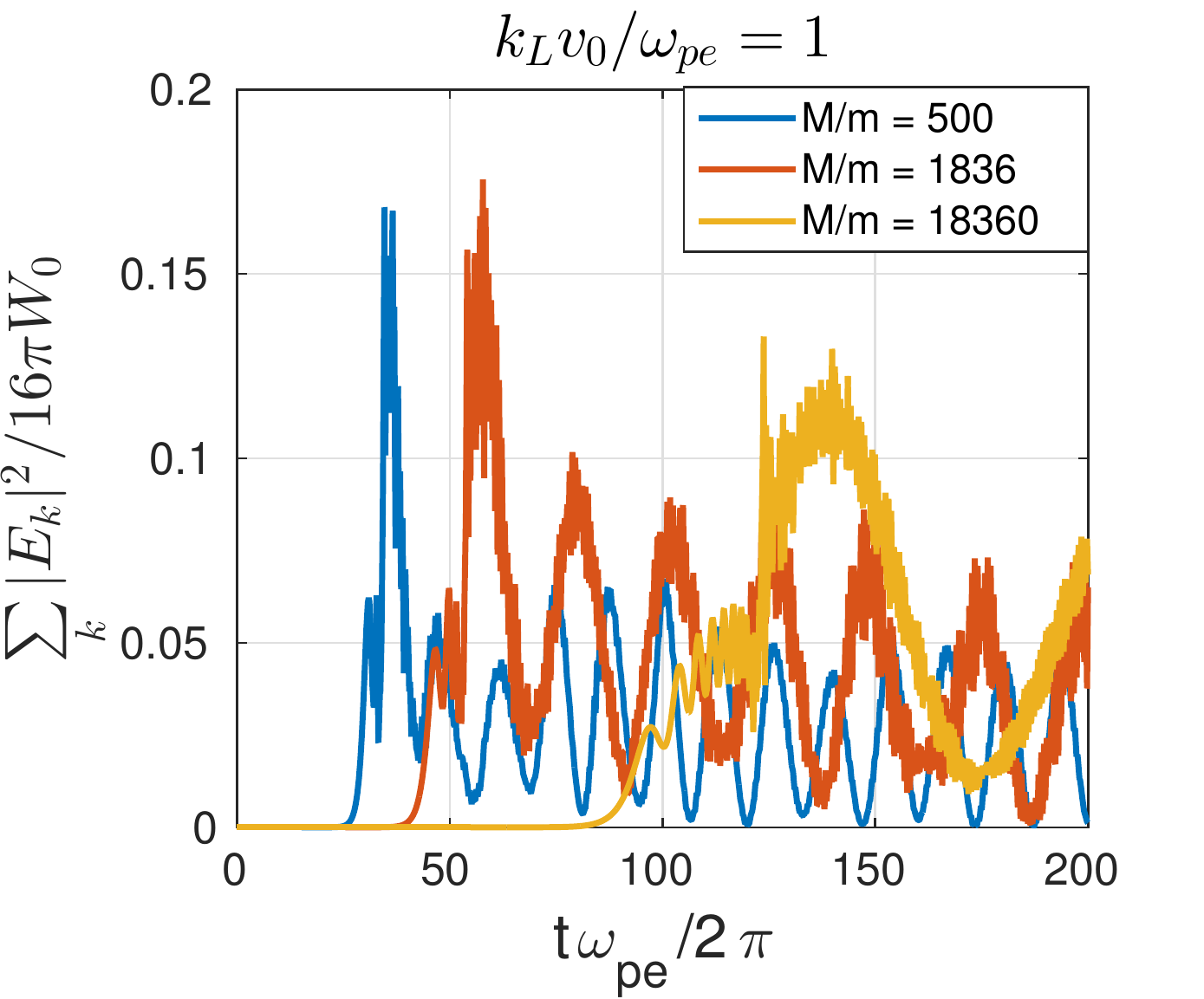}
}
\caption{Time development of ratio of electrostatic energy density to initial kinetic drift energy density with various mass ratio for the initial drift velocities (a) $k_{L}v_{0}/\omega_{pe} \approx 0.33$ and (b) $k_{L}v_{0}/\omega_{pe} \approx 1$.}
\end{figure}
\begin{figure}
\centering
	\includegraphics[scale=0.6]{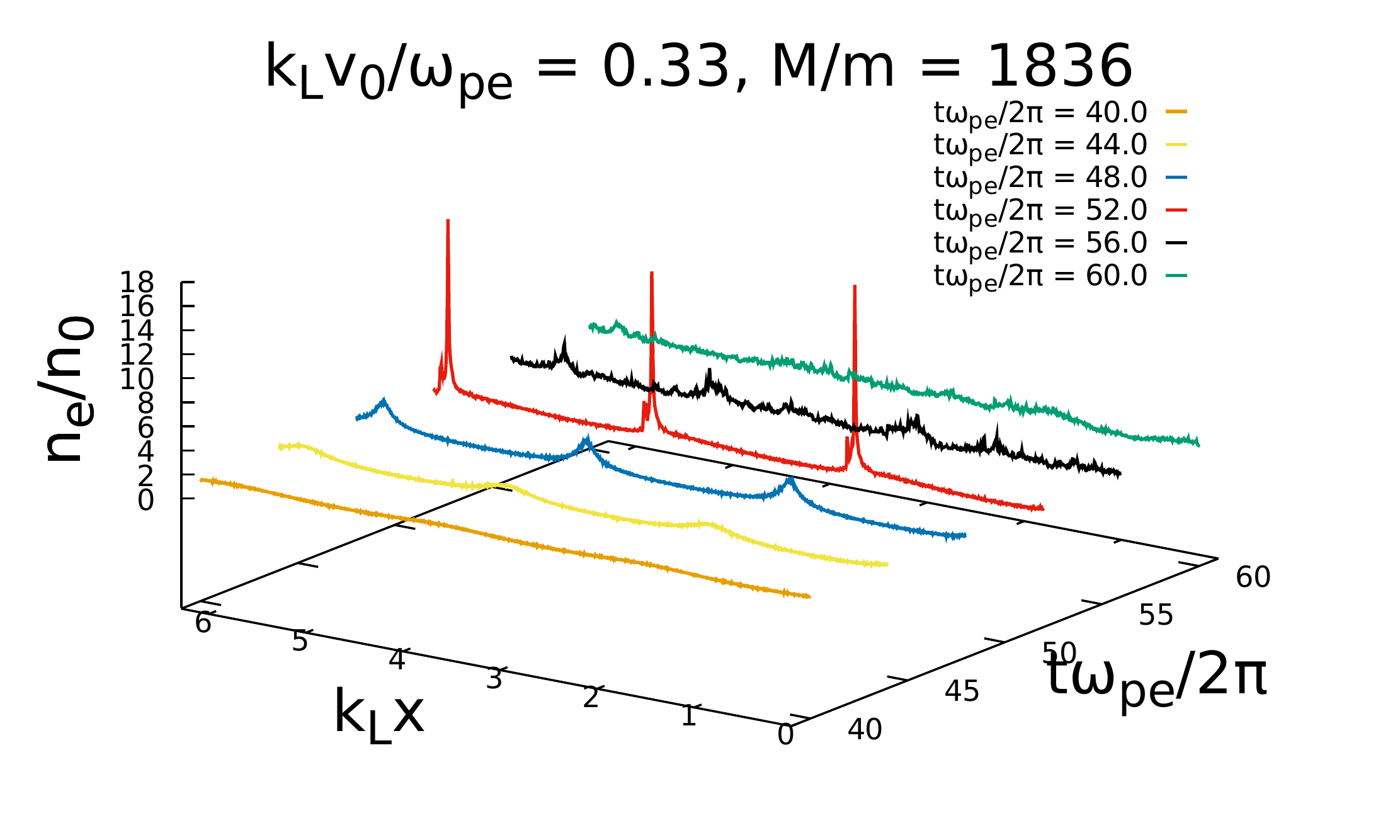}
\caption{Evolution of electron density at different time instances for $k_{L}v_{0}/\omega_{pe} \approx 0.33$ and M/m = 1836.}
\label{fig:ewb}
\end{figure}
\begin{figure}
\centering
	\includegraphics[scale=0.6]{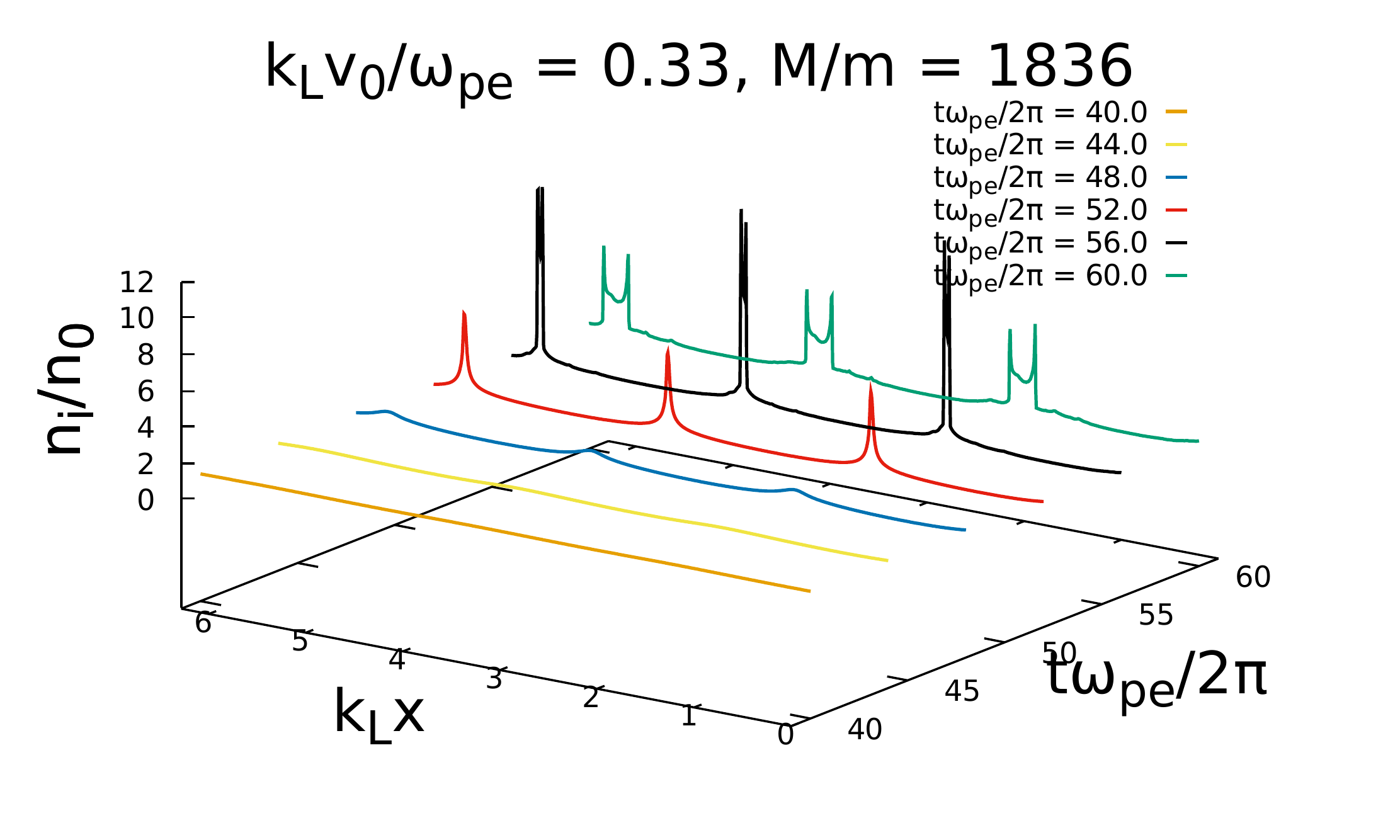}
\caption{Evolution of ion density at different time instances for $k_{L}v_{0}/\omega_{pe} \approx 0.33$ and M/m = 1836.}
\label{fig:iwb}
\end{figure}
%
%\begin{figure} \label{potential}
%\centering
%\includegraphics[scale=0.5]{m_M_1836/k_3/potential_phase_reversal_wpt_45_55.pdf}
%\caption{Phase reversal of electrostatic potential during coherent particle trapping.}
%\end{figure}
%
\begin{figure}
\centering
\subfloat[]{
	\includegraphics[scale=0.13]{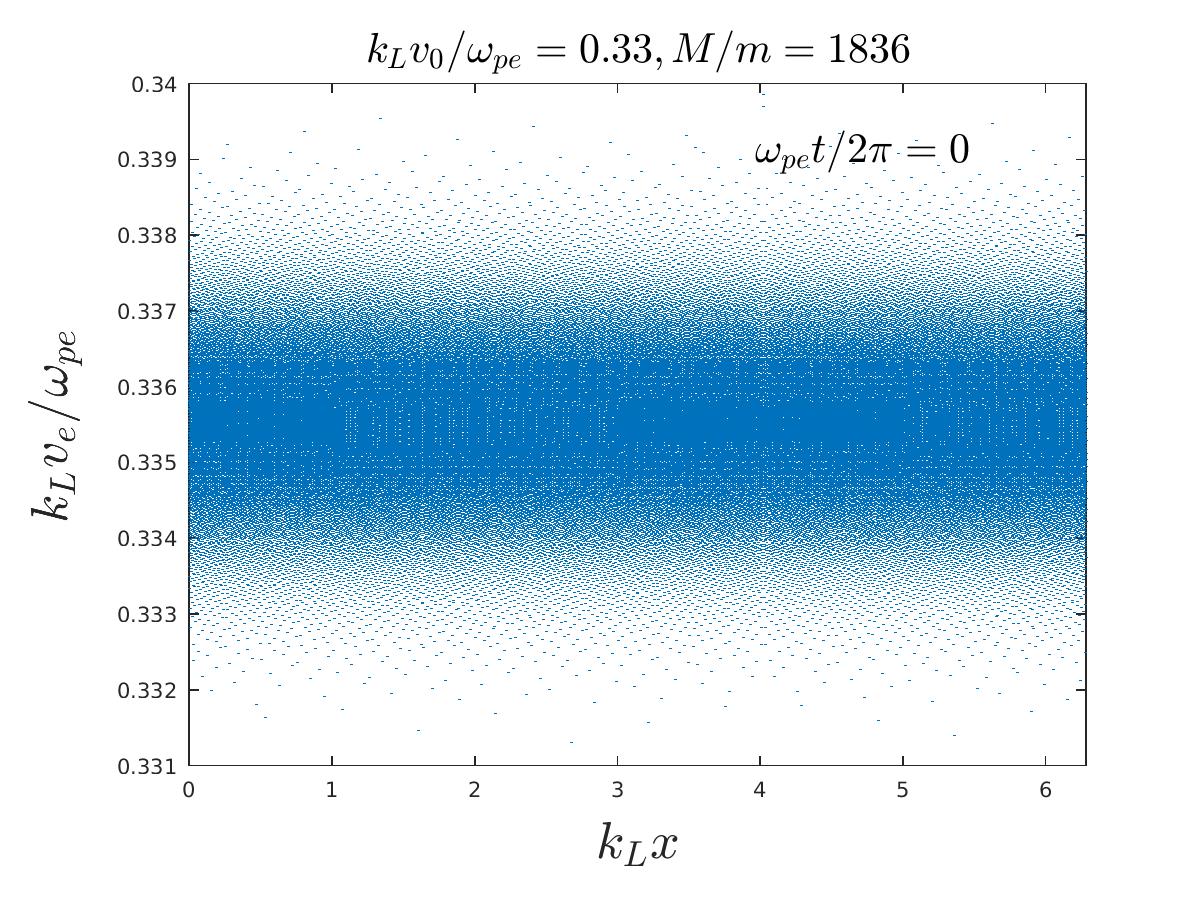}
	\label{fig:eps1}
}
\subfloat[]{
	\includegraphics[scale=0.13]{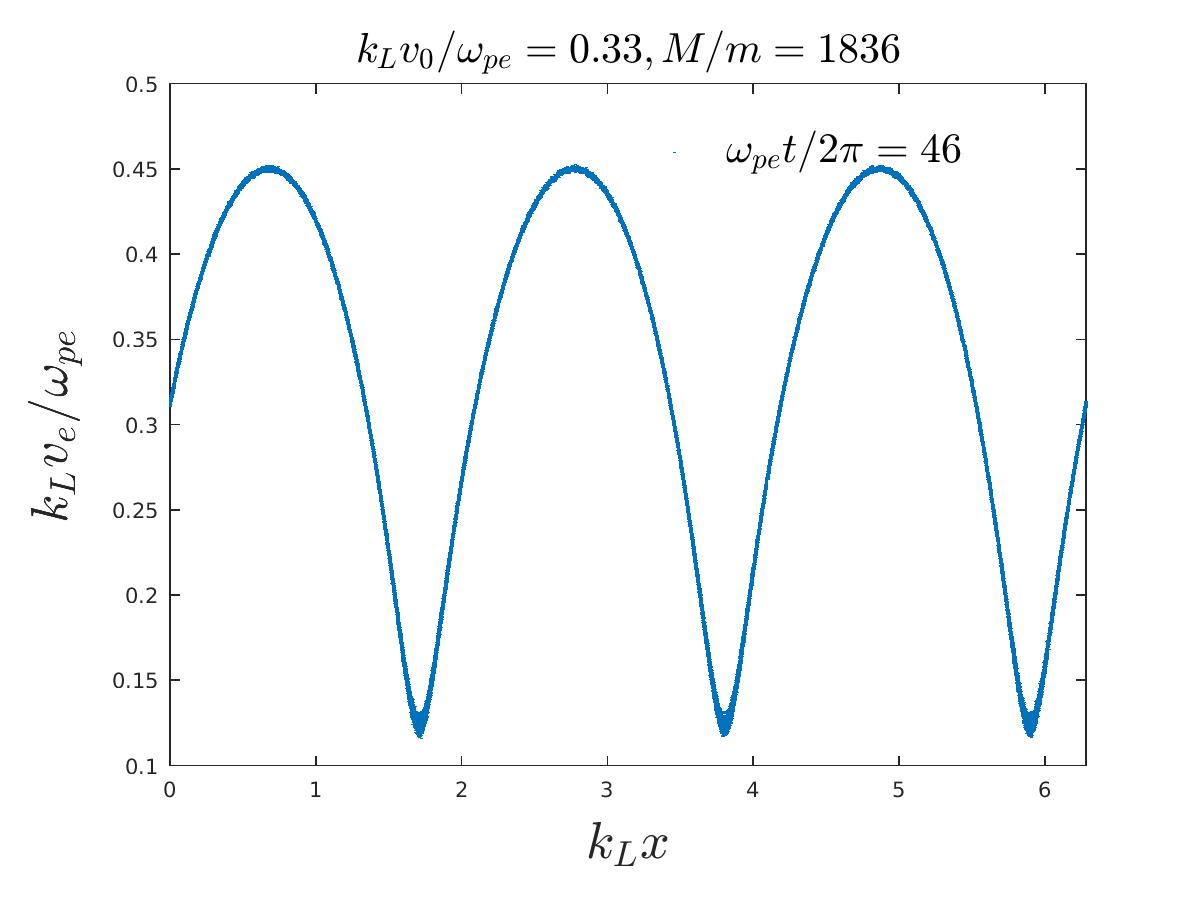}
    \label{fig:eps2}
}
\subfloat[]{
	\includegraphics[scale=0.13]{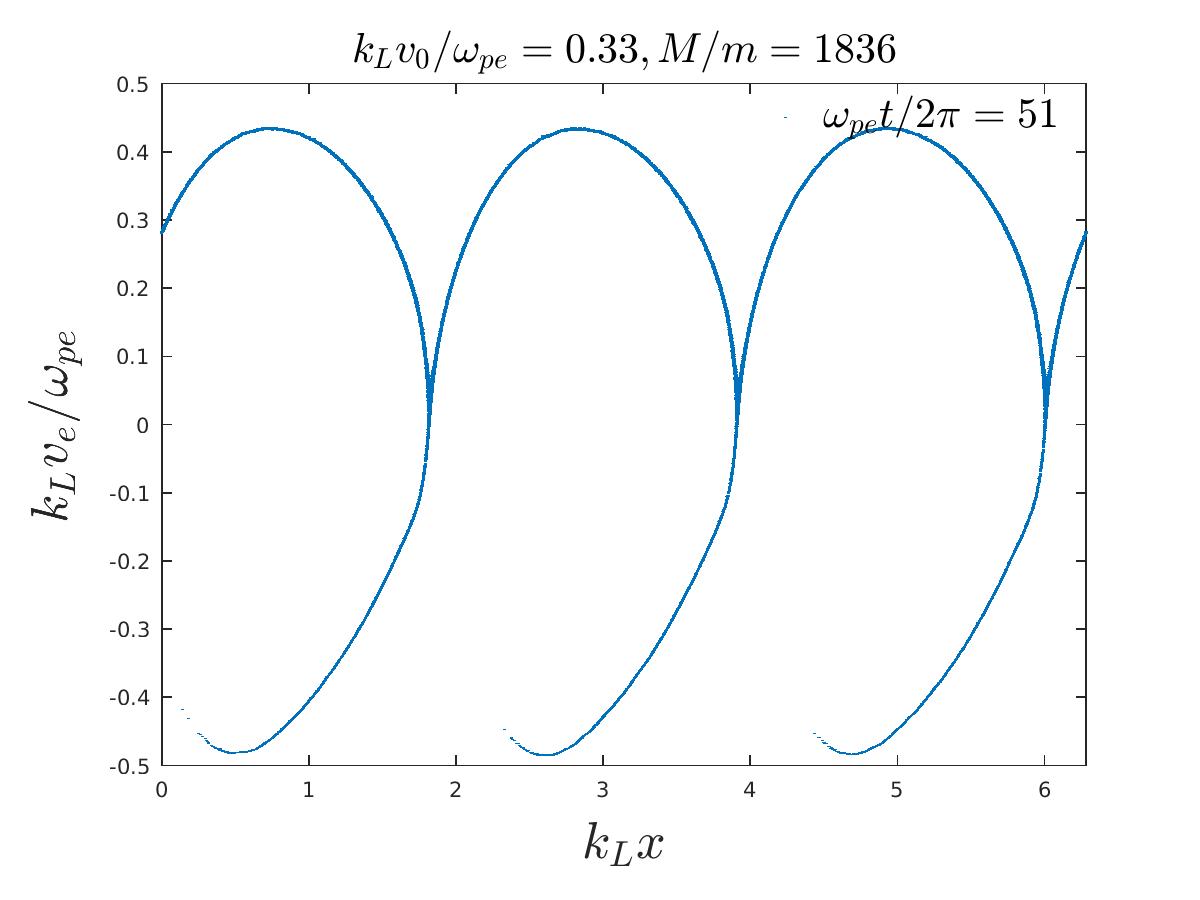}
	\label{fig:eps3}
}\\
\subfloat[]{
	\includegraphics[scale=0.13]{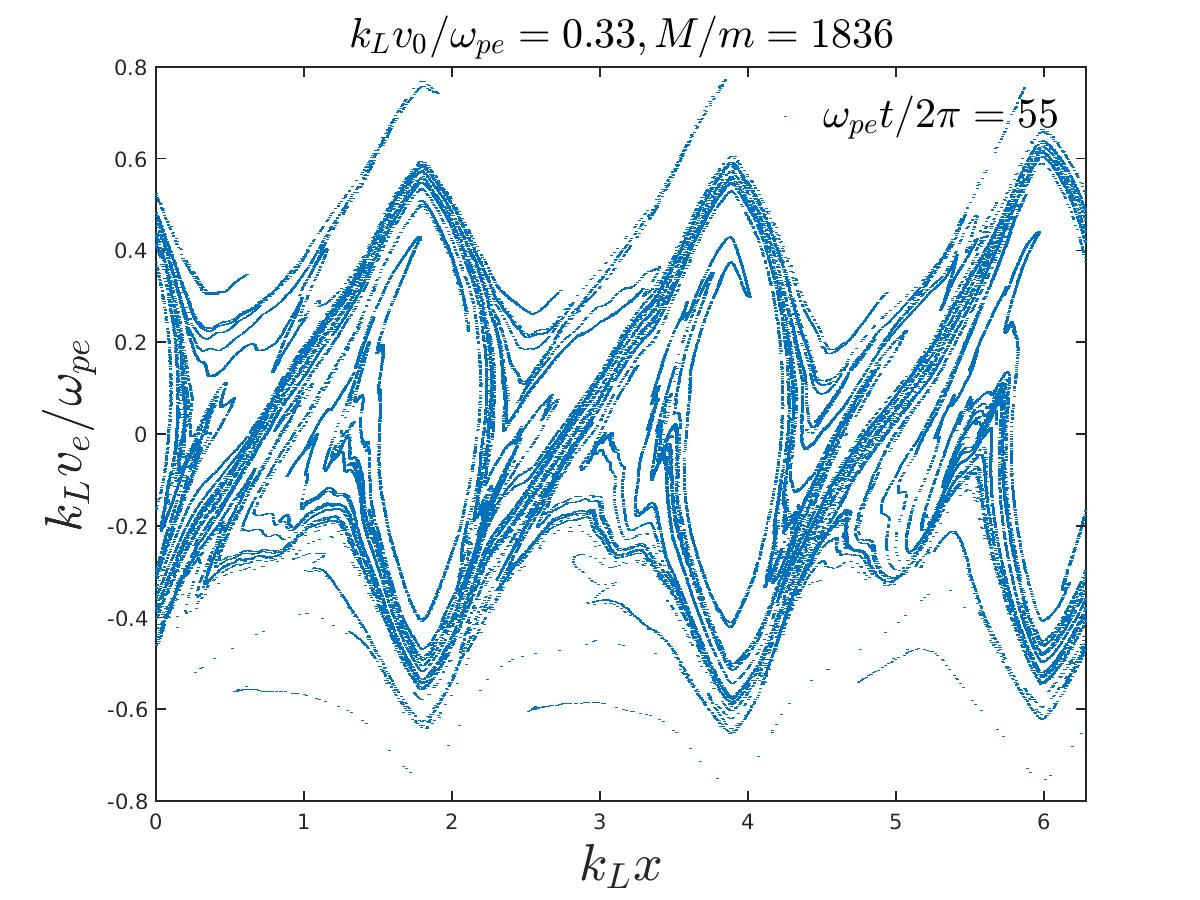}
	\label{fig:eps4}
}
\subfloat[]{
	\includegraphics[scale=0.13]{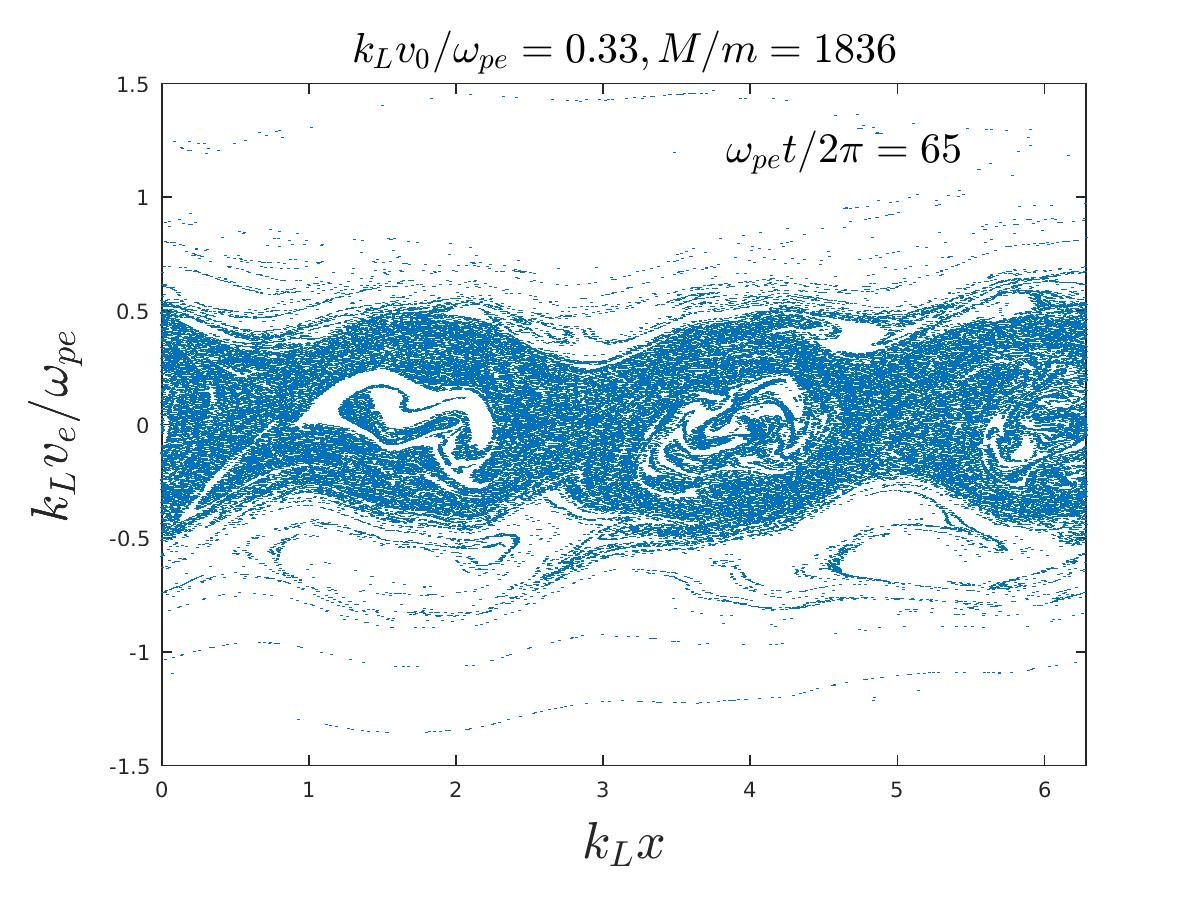}
	\label{fig:eps5}
}
\subfloat[]{
	\includegraphics[scale=0.13]{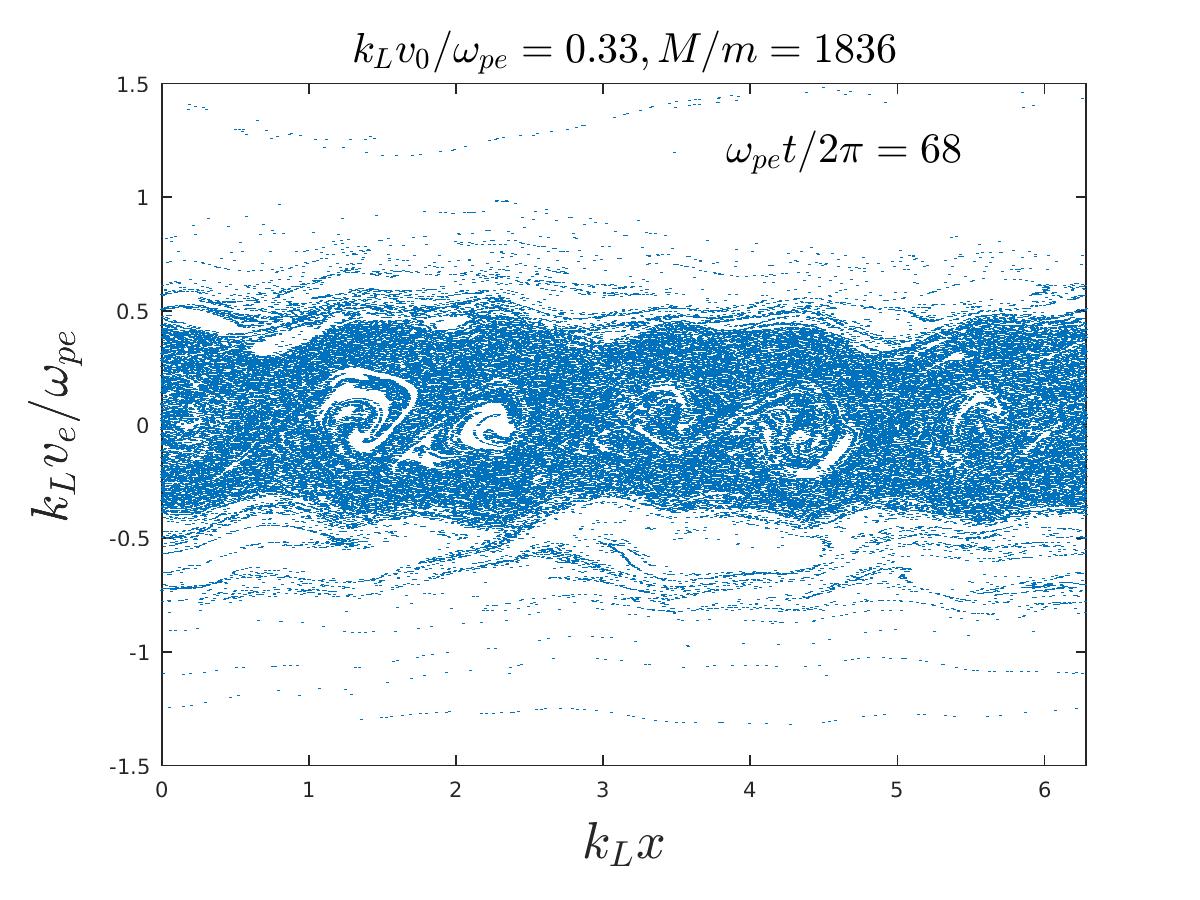}
	\label{fig:eps5}
}\\
\subfloat[]{
	\includegraphics[scale=0.13]{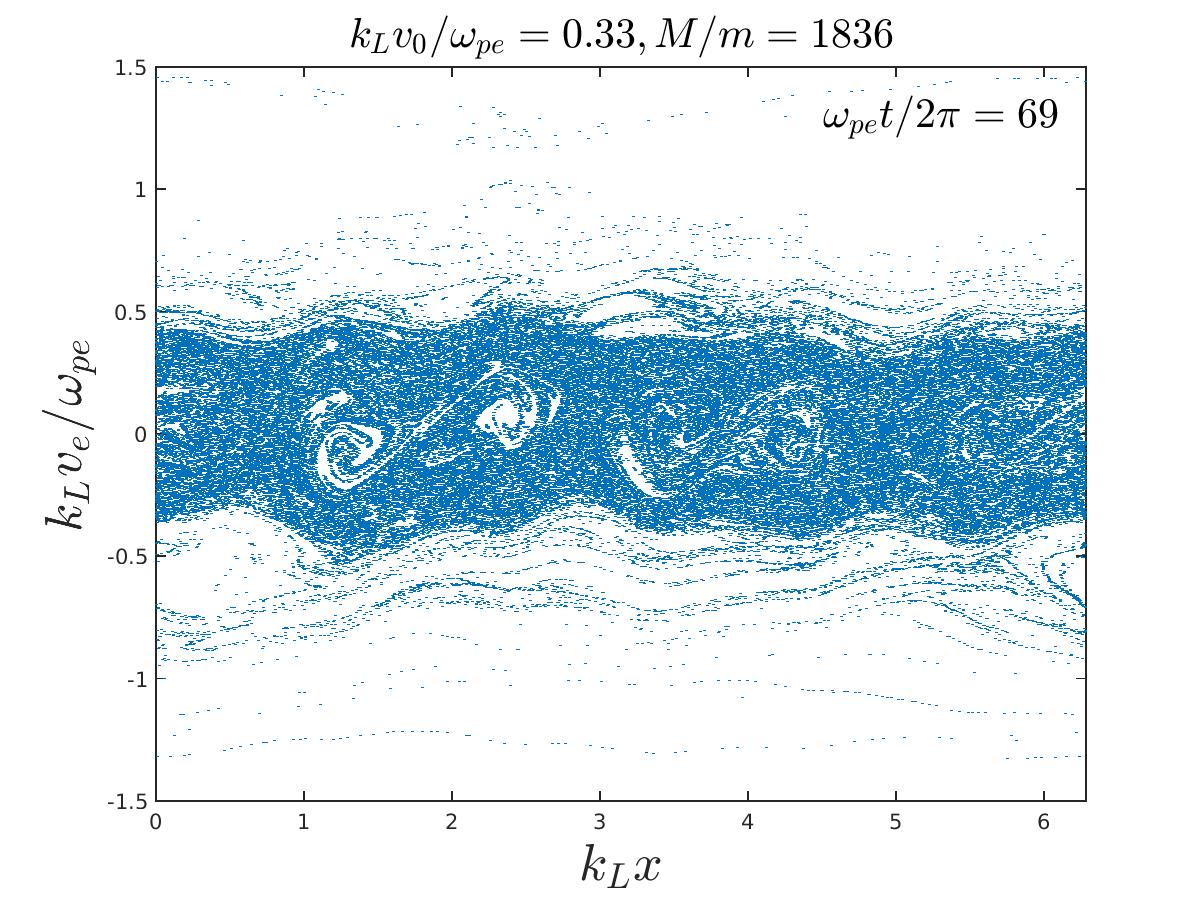}
	\label{fig:eps5}
}
\subfloat[]{
	\includegraphics[scale=0.13]{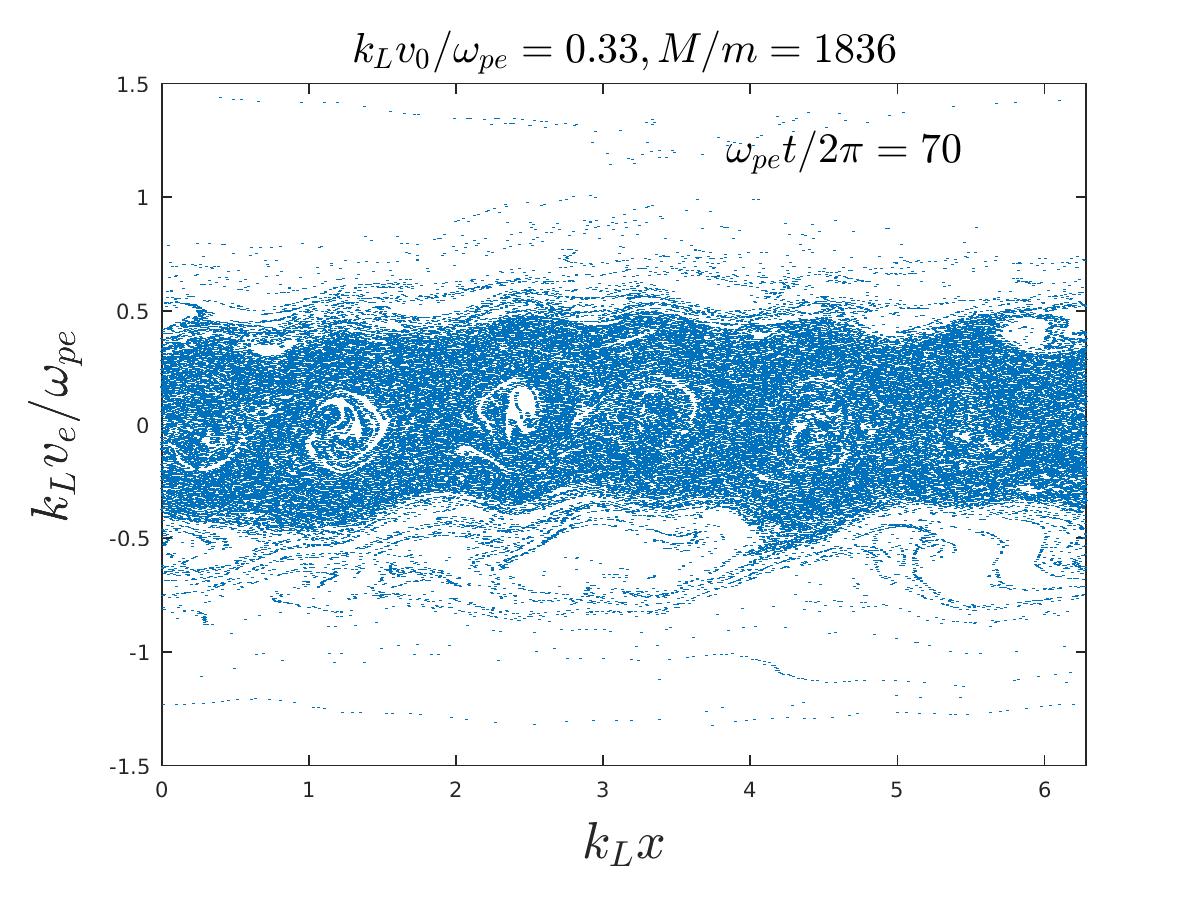}
	\label{fig:eps6}
}
\subfloat[]{
	\includegraphics[scale=0.13]{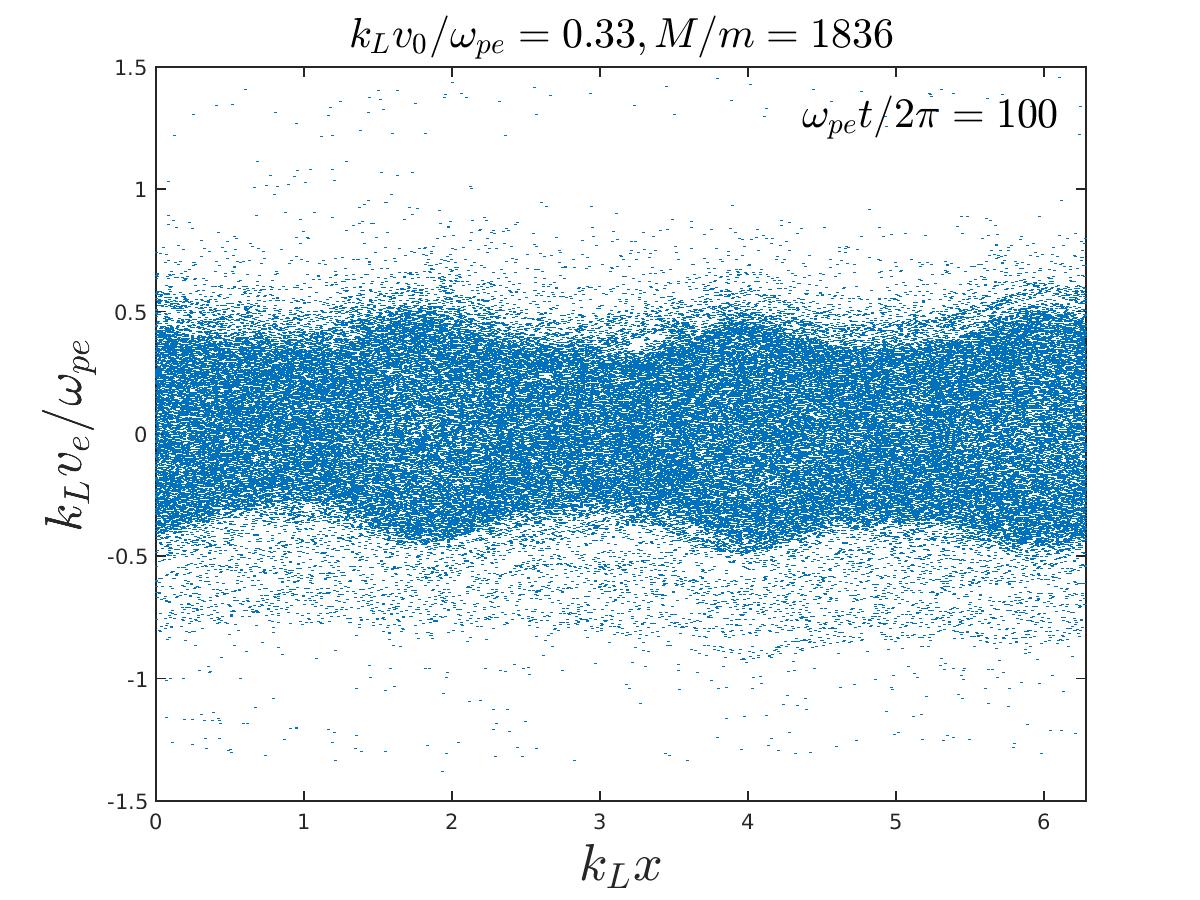}
	\label{fig:eps6}
}
\caption{Evolution of electron phase space at the different stages of simulation for $k_{L}v_{0}/\omega_{pe} \approx 0.33$ and M/m = 1836.}
\label{fig:electron phase space}
\end{figure}
\begin{figure} \label{potential}
\centering
\includegraphics[scale=0.6]{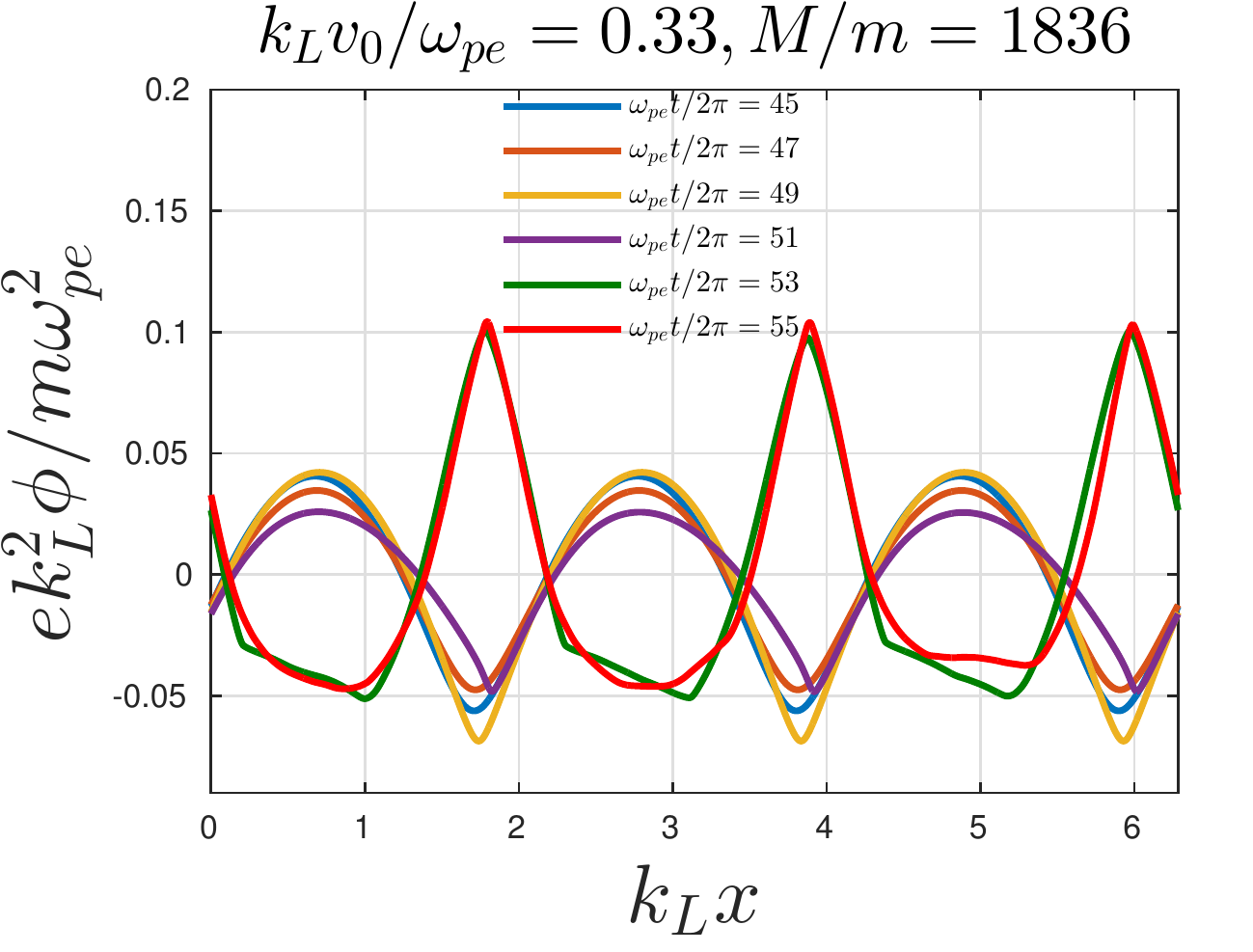}
\caption{Phase reversal of electrostatic potential during particle trapping for $k_{L}v_{0}/\omega_{pe} \approx 0.33$ and M/m = 1836.}
\end{figure}
\begin{figure}
\centering
\subfloat{
\includegraphics[scale=0.3]{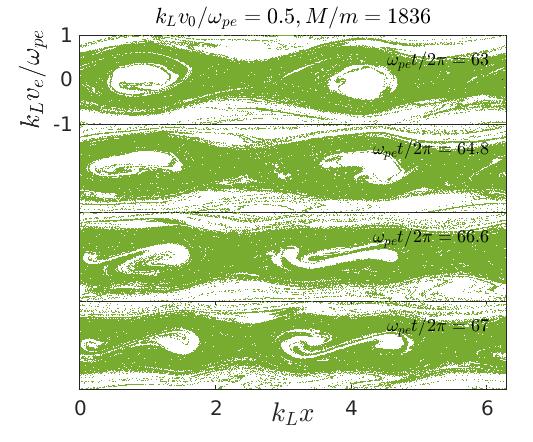}
\includegraphics[scale=0.3]{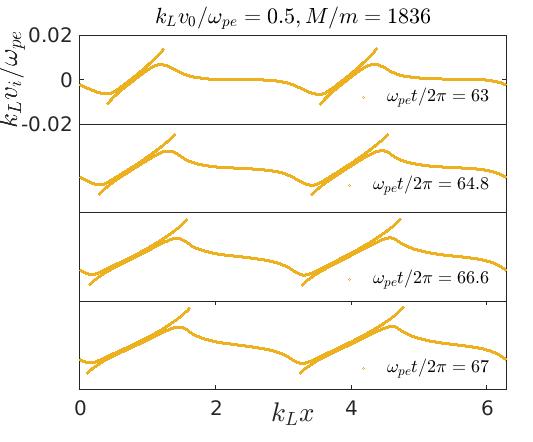}
\includegraphics[scale=0.35]{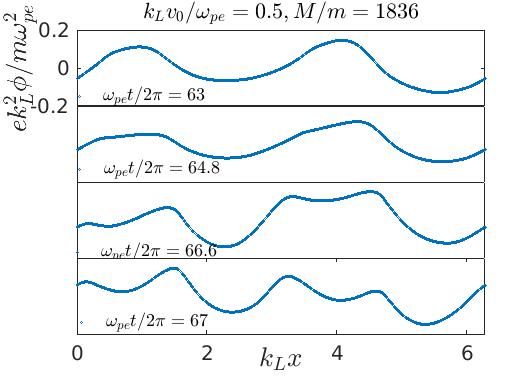}
}\\
\subfloat{
\includegraphics[scale=0.3]{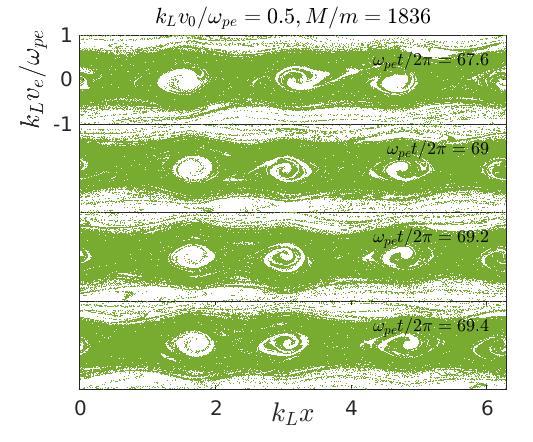}
\includegraphics[scale=0.3]{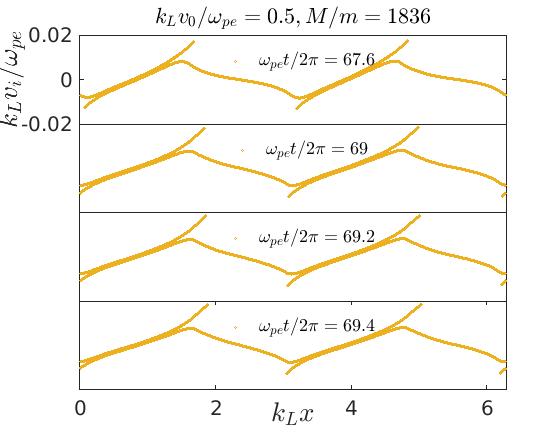}
\includegraphics[scale=0.35]{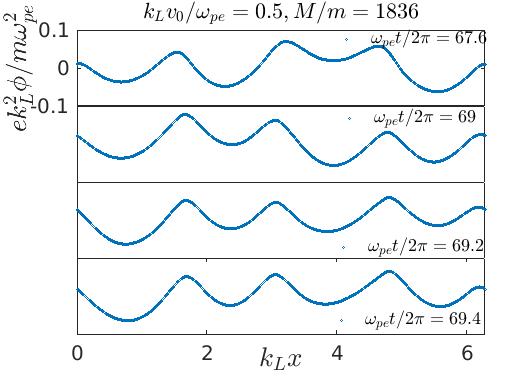}
}
\label{fig:chs}
\caption{Breaking of electron hole and generation of CHS for $k_{L}v_{0}/\omega_{pe} = 0.5$ and M/m = 1836.}

\end{figure}
\begin{figure}
\centering
\subfloat[]{ \label{fig:water bag phase space}
\includegraphics[scale=0.45]{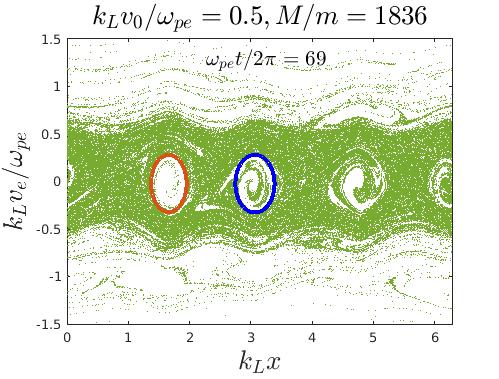}
}
\subfloat[]{ \label{fig:water bag distribution}
\includegraphics[scale=0.57]{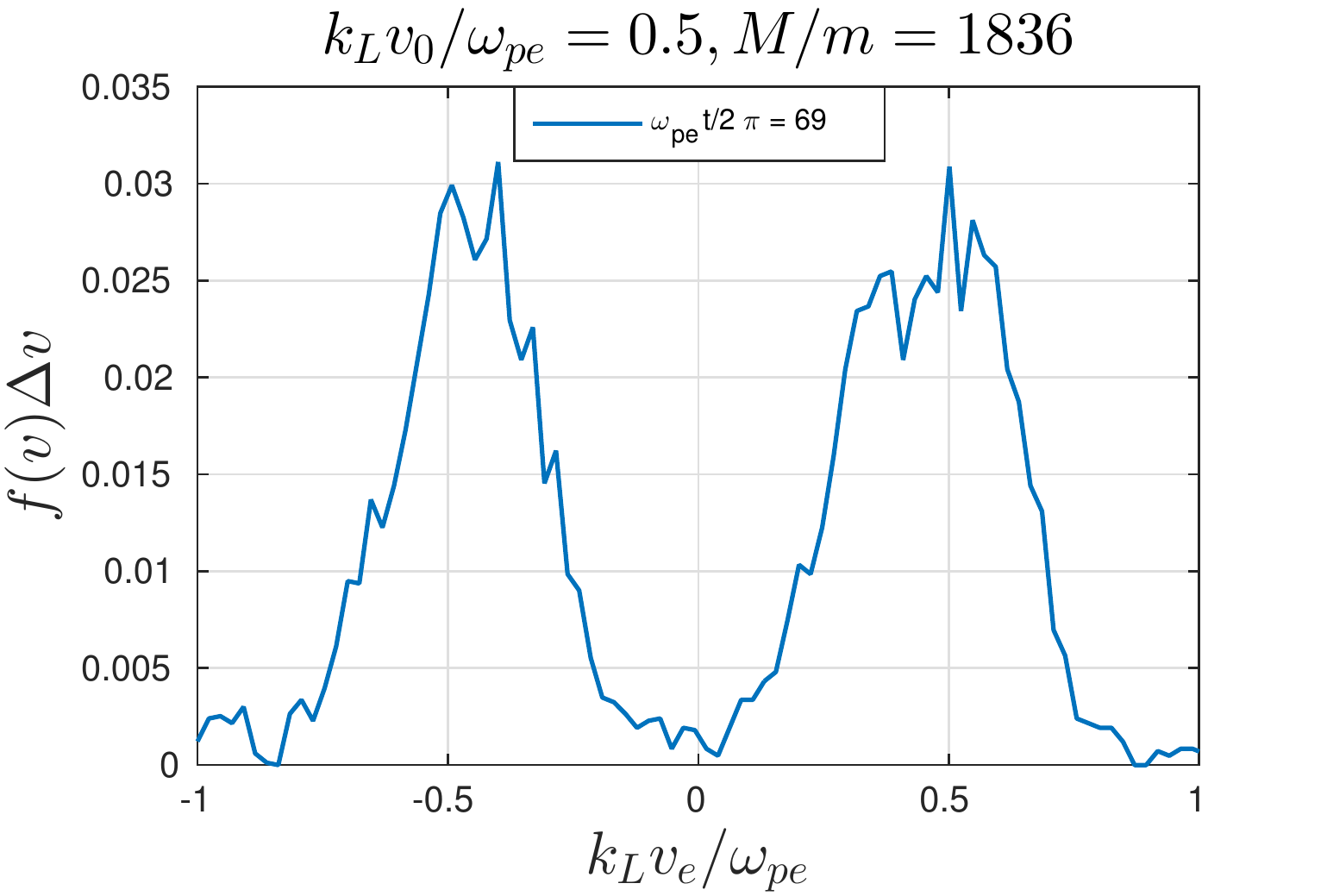}
}
\caption{Electron distribution function for a CHS which closely resembles to the water bag distribution.}
\end{figure}
\begin{figure}
\centering
\label{fig:hole area 1836}
\includegraphics[scale=0.45]{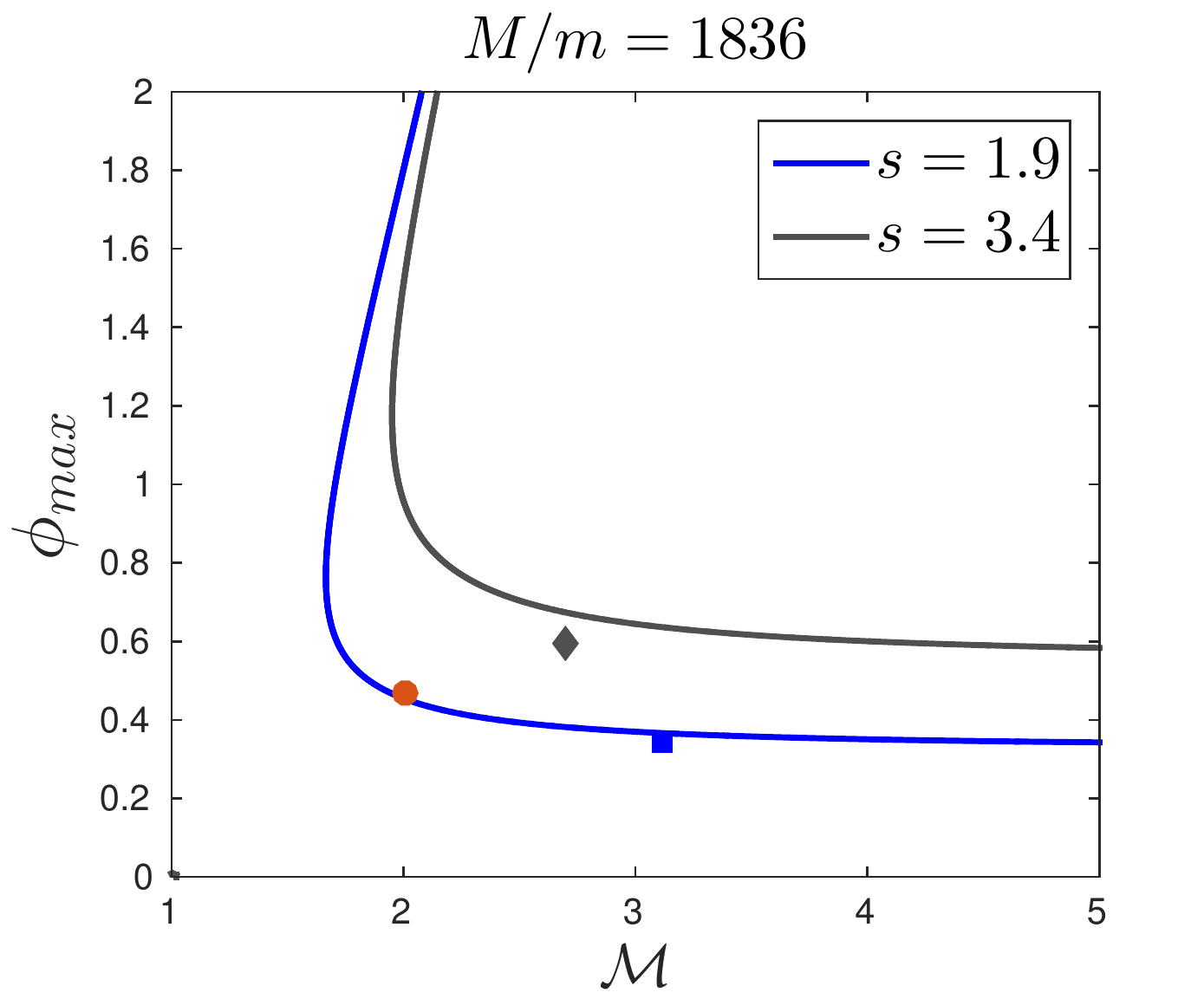}
\caption{Theoretical $\mathcal{M} - \phi_{max}$ curve for the mass ratio  M/m = 1836. Lines show theoretical relation for a fixed area of the CHS while dots are taken from simulation. }
\end{figure}
\begin{figure}
\centering
\label{fig:hist0_55_100}
\includegraphics[scale=0.4]{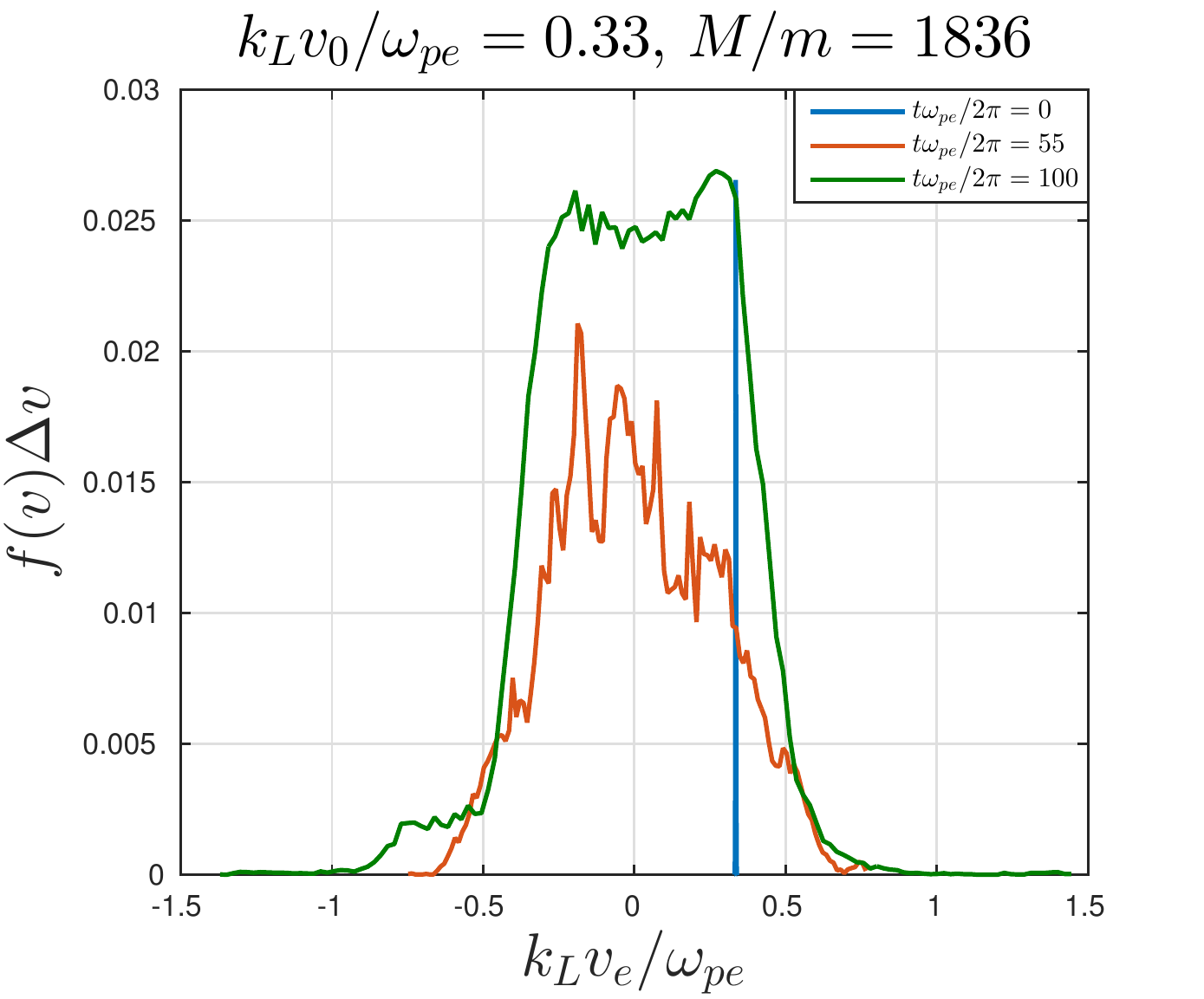}
\caption{Evolution of electron distribution function at the time $\omega_{pe}t/2\pi \approx$ 0, 55 and 100.}
\label{fig:electron distribution function}
\end{figure}
%
%

%%%%%%%%%%%%%%%%%%%%%%%%%%%%%%%%%%%%%%%%%%%%%%%%%%%%%%%%%%%%%%%%%%%%%%%%%%%%%%%%%%%%%%%%%%%%%%%%%%%%%%%%%%%%%%%
\end{document}